\documentclass[12pt,preprint]{aastex}

\altaffiltext{1}{Laboratoire AIM, CEA/DSM-CNRS-Universit\'e Paris Diderot, DAPNIA/Service d'Astrophysique, B\^at. 709, CEA-Saclay, F-91191 Gif-sur-Yvette C\'edex, France ; \textit{Benjamin.magnelli@cea.fr}}
\altaffiltext{2}{Spitzer Science Center, California Institute of Technology, Pasadena, CA 91125, USA; \textit{rchary@caltech.edu}}
\altaffiltext{3}{{\it Spitzer} Fellow; National Optical Astronomy Observatory, 950 North Cherry Avenue, Tucson, AZ 85719, U.S.A.}
\altaffiltext{4}{University of British Columbia, Vancouver, BC V6T 1Z1, Canada}
\altaffiltext{5}{Institute of Astronomy, University of Hawaii, Honolulu, HI, 96822, USA}
\altaffiltext{6}{Canada-France-Hawaii Telescope, Kamuela, HI, 96743, USA}
\altaffiltext{7}{National Optical Astronomy Observatory, Tucson, AZ 85719}
\shortauthors{Magnelli et al.}
\shorttitle{The $3.3\,\mu$m PAH Feature in $z>0.6$ Star-Forming Galaxies}

\begin{document}

\title{IRAC Excess in Distant Star-Forming Galaxies: Tentative Evidence for the 3.3$\mu$m Polycyclic Aromatic Hydrocarbon Feature ?}
\author{B. Magnelli\altaffilmark{1,2}, R.~R. Chary\altaffilmark{2}, A. Pope\altaffilmark{3,4}, D. Elbaz\altaffilmark{1}, G. Morrison\altaffilmark{5,6} \& M. Dickinson\altaffilmark{7}}
\keywords{Galaxies: evolution - Infrared: galaxies - Galaxies: starburst}


\begin{abstract}
We present evidence for the existence of an IRAC excess in the spectral energy distribution (SED) of 5 galaxies at $0.6<z<0.9$ and 1 galaxy at $z=1.7$.
These 6 galaxies, located in the Great Observatories Origins Deep Survey field (GOODS-N), are star forming since they present strong $6.2,\ 7.7$, and,$\,11.3\,\mu$m polycyclic aromatic hydrocarbon (PAH) lines in their \textit{Spitzer} IRS mid-infrared spectra.
We use a library of templates computed with PEGASE.2 to fit their multiwavelength photometry and derive their stellar continuum.
Subtraction of the stellar continuum enables us to detect in 5 galaxies a significant excess in the IRAC band pass where the
 $3.3\,\mu$m PAH is expected  (i.e IRAC $5.8\,\mu$m for the range of redshifts considered here).
We then assess if the physical origin of the IRAC excess is due to an obscured active galactic nucleus (AGN) or warm dust emission.
For one galaxy evidence of an obscured AGN is found, while the remaining four do not exhibit any significant AGN activity.
Possible contamination by warm dust continuum of unknown origin as found in the Galactic diffuse emission is discussed. 
The properties of such a continuum would have to be different from the local Universe
to explain the measured IRAC excess, but we cannot definitively rule out this possibility until its origin is understood.
Assuming that the IRAC excess is dominated by the $3.3\,\mu$m PAH feature, we find good agreement with the observed $11.3\,\mu$m PAH line flux arising from the same C-H bending and stretching modes, consistent with model expectations.
Finally, the IRAC excess appears to be correlated with the star-formation rate in the galaxies.
Hence it could provide a powerful diagnostic for measuring dusty star formation in $z>3$ galaxies once the mid-infrared spectroscopic capabilities of the \textit{James Webb Space Telescope} become available.\\ \\
\end{abstract}
\maketitle
\section{Introduction}
\indent{
Measuring the star formation history of galaxies as a function of redshift enables the build up of stellar mass in the Universe to be constrained.
Although there are systematic uncertainties between different star formation tracers, results from different studies (UV, IR) seem to converge on a 
flat or a gradual decrease in the 
star formation rate (SFR) between $z=3$ and $z=1$  followed by a steeper decline between $z=1$ and $z=0$ \citep[e.g.][]{Schiminovich2005,chary_2001}.\\
}
\indent{
At high redshift ($z>3$), evolution of the SFR still remains uncertain, primarily due to the poorly determined dust extinction correction.
At these redshifts the SFR in galaxies is estimated using the UV luminosity which is strongly affected by dust extinction \citep{steidel_1999}.
The extinction correction can be calculated using the UV slope \citep{meurer_1999, adelberger_2000} but this technique is affected by several limitations.
The UV slope can be influenced by the presence of an evolved stellar population and therefore overestimates the extinction correction.
Moreover it is well known that local galaxies harboring strong dusty star formation are opaque to UV radiation \citep[e.g.][]{buat_2005}.
This implies that the UV luminosity may not be a reliable tracer of the SFR in certain galaxies and that mid and far-infrared tracers which are correlated with dust emission are required.
The strongest dust emission features, such as those arising between $6-12$\,$\mu$m in polycyclic aromatic hydrocarbon (PAH) molecules, are redshifted out of the mid-infrared passband at $z>3$. As a result, even deep $24\,\mu$m observations are insensitive to dusty star formation at these redshifts.
Hence, a complete understanding of dust correction will require deep observations in the far-infrared/submillimeter regime or require accurate calibration of near-infrared tracers of dusty star formation.\\
}
\indent{
The \textit{James Webb Space Telescope} (JWST) will obtain spectra between $5-27 \mu$m and thereby detect the redshifted 3.3$\mu$m PAH feature from galaxies at $z>3$.
The success of previous studies using monochromatic PAH luminosities as dusty SFR tracers \citep[e.g.][]{chary_2001,takeuchi_2005, brandl_2006} motivates the calibration of this PAH emission feature as a SFR indicator.\\
}
\indent{Some studies have already investigated the possibility of using the $3.3\,\mu$m PAH feature as a SFR tracer and have revealed the presence of a correlation between $L_{3.3\,\mu m}$ and SFR in local galaxies \citep{mouri_1990,imanishi_2002,imanishi_2006}. 
However the existence of this 3.3$\mu$m PAH feature at higher redshifts needs to be confirmed and its line strength calibrated to the true SFR of galaxies.\\
}
 \indent{
 In this paper we analyze the multi-wavelength properties of 6 galaxies at $z>0.5$ and assess the evidence for the presence of the 3.3$\mu$m PAH feature.
The galaxies display strong $6.2, 7.7$, and$\,11.3\,\mu$m PAH emission in their \textit{Spitzer} mid-infrared spectra and are hence expected to also show 3.3$\mu$m PAH emission.\\}
\indent{ 
The sample of galaxies, situated in the Great Observatories Origin Deep Survey-North \citep[GOODS-N,][]{dickinson_goods} field, has been observed in the optical by the \textit{Hubble Space Telescope} (HST) and in the infrared by the \textit{Spitzer Space Telescope}.
The multiwavelength photometry is fitted with spectral energy distributions (SEDs) from the stellar population synthesis model PEGASE.2 \citep{fioc_1997}.
The residual emission in the IRAC passband, where the 3.3$\mu$m PAH signature should be present, is analyzed to determine its origin.
We first investigate the possibility that this excess originates from an obscured active galactic nucleus (AGN) using the SED of NGC 1068.
Then we assess the possibility that this excess originates from warm dust emission i.e the 3.3 $\mu$m PAH line emission and/or a warm dust continuum.
To test the hypothesis that the IRAC excess is dominated by the $3.3\,\mu$m PAH line flux we compare the derived line flux to the 11.3 $\mu$m PAH flux measured in the IRS spectra.
Since models predict that the $3.3\,\mu$m and $ 11.3 \mu$m PAH lines originate from C-H modes \citep[][]{li_2001}, their line fluxes should be correlated.\\}
\indent{The layout of the paper is as follows:  The sample is presented in Section \ref{sec: samples}, the SED fits with PEGASE.2 and the determination of the IRAC excess is presented in Section \ref{sec: data analysis}. Section \ref{sec:Origin} discusses the origin of the IRAC excess which at these wavelengths can be due to hot dust emission from an obscured AGN (Section \ref{sec:AGN}), free-free and/or recombination line emission (Section \ref{sec:freefree}) or finally to the $3.3\,\mu$m PAH broad emission line and/or a warm dust continuum (Section \ref{sec:cont} and \ref{sec:PAH}).
Our conclusions are summarized in Section \ref{sec: conclusion}. \\ \\}
\indent{
Throughout this paper we will use a cosmology with $H_{0}=71\ km\ s^{-1}\ Mpc^{-1},\Omega_{\Lambda}=0.73 ,\Omega_{M}=0.27$.\\
}
\section{Sample selection}
\label{sec: samples}
\begin{figure}
 \plottwo{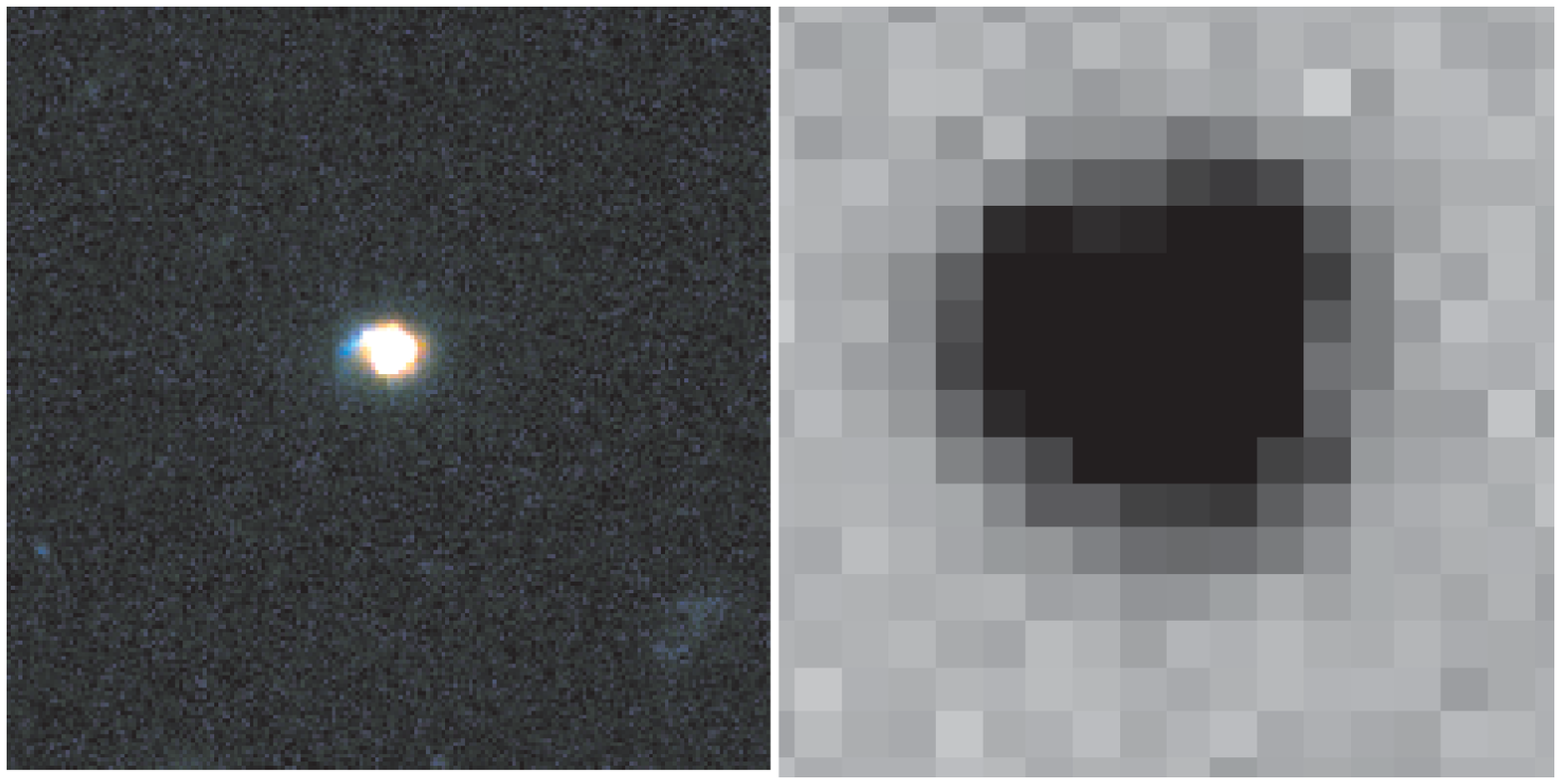}{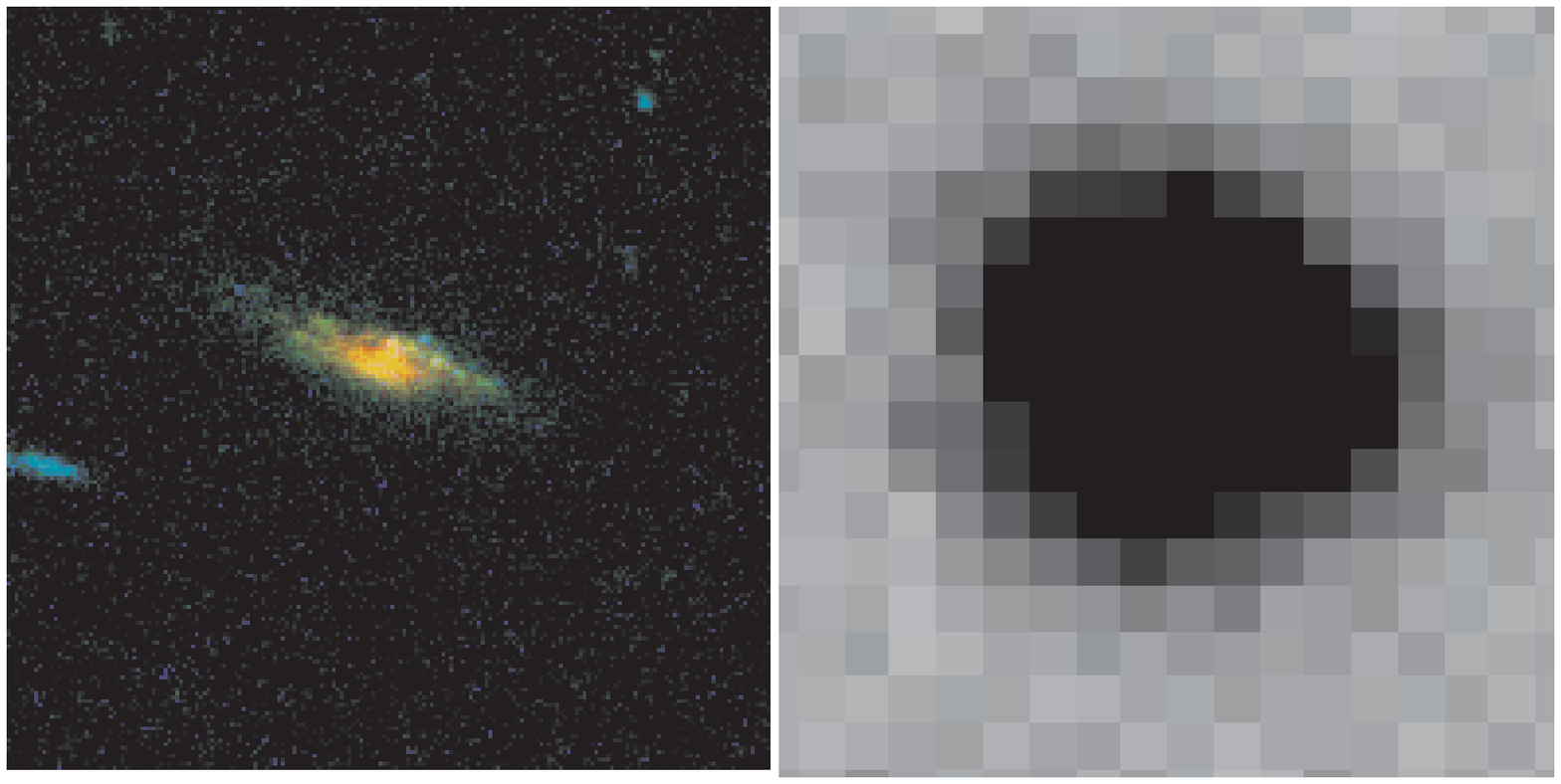}\\
 \plottwo{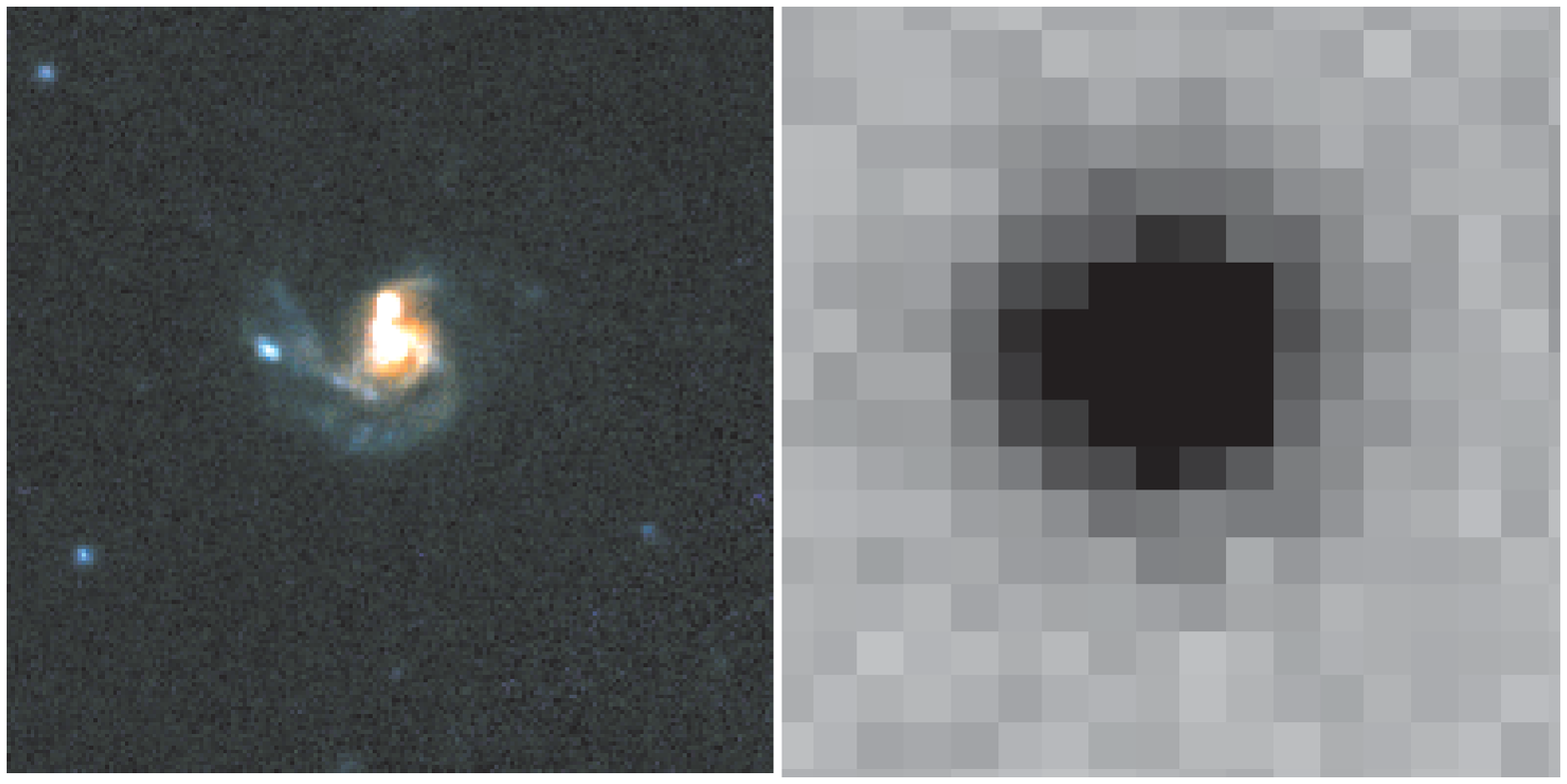}{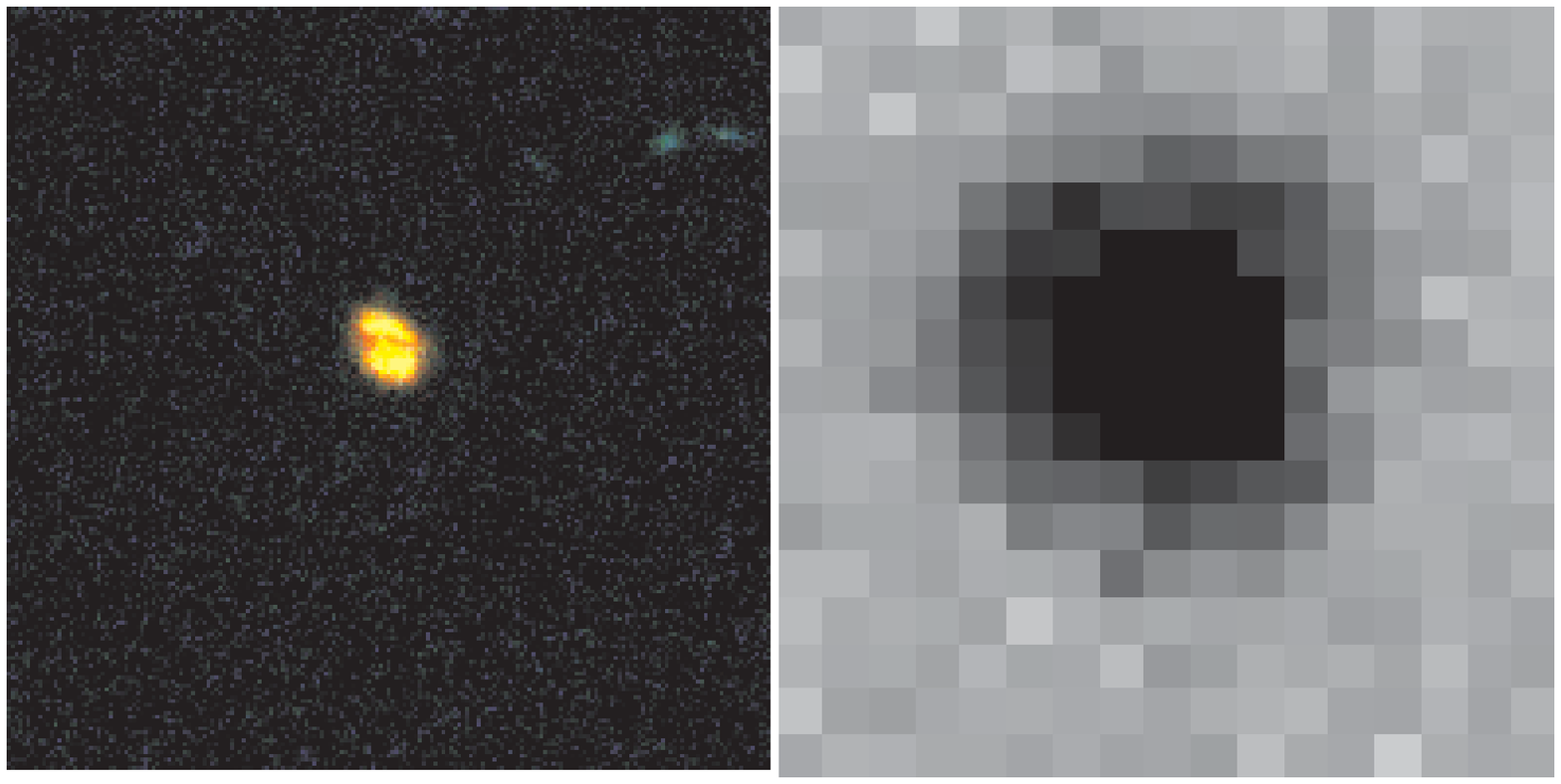}\\
 \plottwo{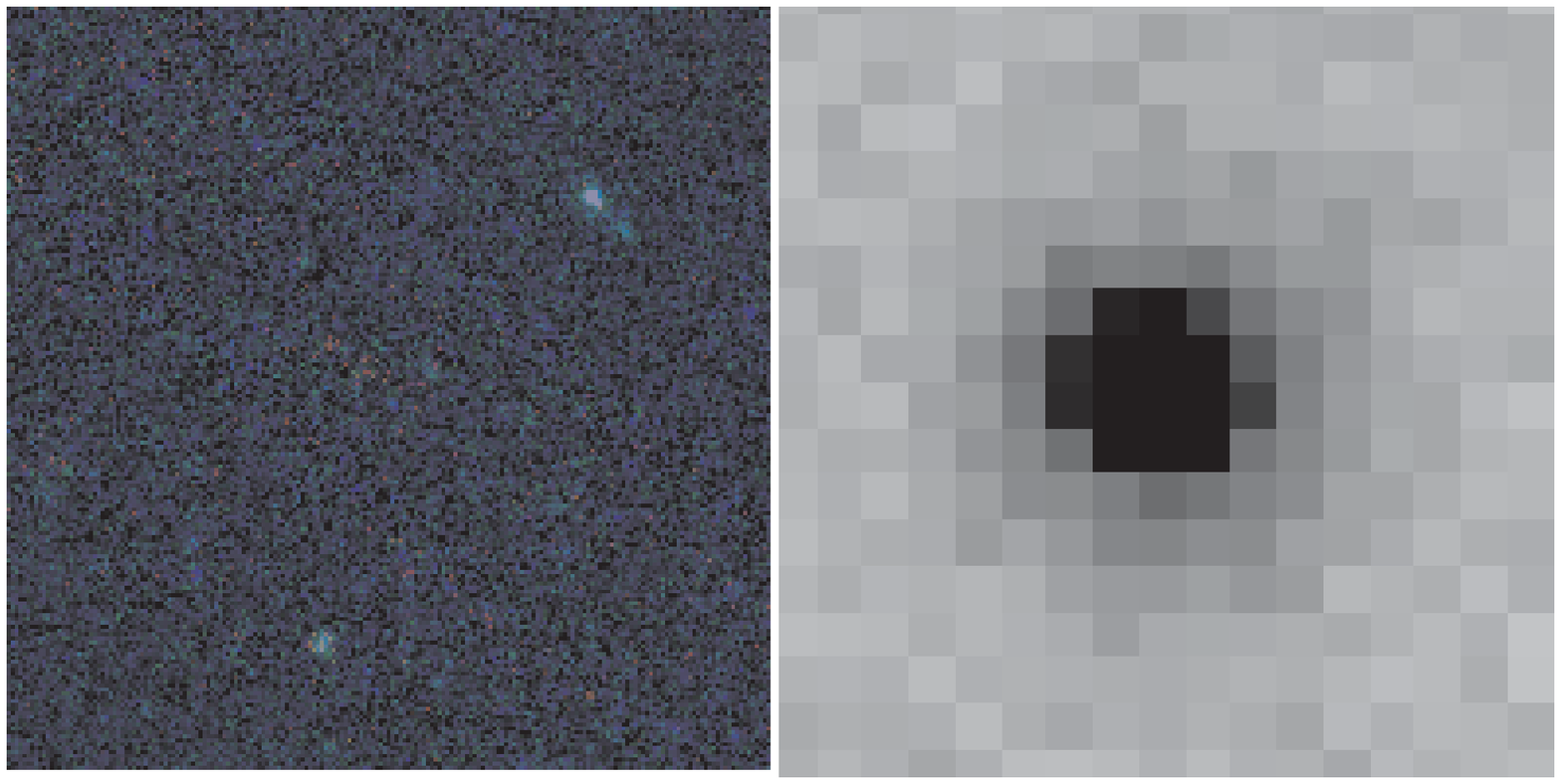}{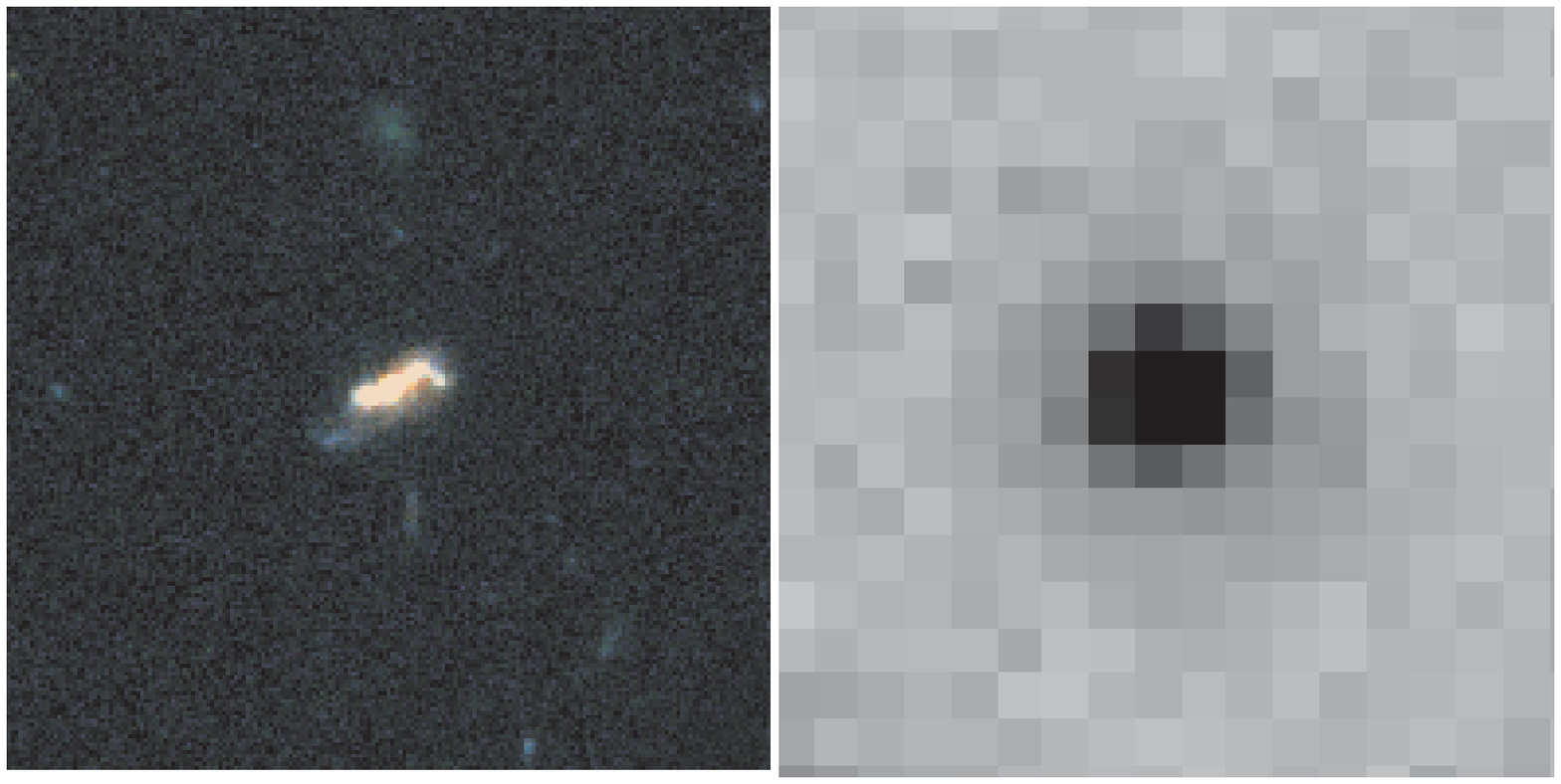}\\
\caption{\label{fig:imagette} \footnotesize{From left to right and from top to bottom each panel shows optical and NIR
  images of  MIPS4, MIPS5, MIPS6, MIPS7, MIPS3419, and MIPS5581
  respectively. For each galaxy, we
  show on the left a color composite $BVz$ image from HST and on the right IRAC 5.8
  $\mu$m from \textit{Spitzer}. Each image spans a
 $10\arcsec\times10\arcsec$ area.}}
\end{figure}

\indent{
The 3.3 $\mu$m and the 11.3 $\mu  m$ PAH lines arise from stretching and out-of-plane bending modes of the C-H bond, respectively \citep{duley_1981,leger_1984,li_2001}.
In order to obtain the best constraints on the $3.3\,\mu$m PAH emission we consider bright 24$\mu$m sources in the GOODS-N field  which have spectral coverage of the 11.3$\,\mu  m$ PAH feature.
We selected a sample of 21 galaxies with mid-infrared spectra from Infrared Spectrograph (IRS; Houck et al. 2004) observations in \textit{Spitzer} programs GO2-20456 (PI: Chary) and 262 (PI: Helou).
Data reduction includes cleaning rogue pixels, removing latent charge build-up, removing the sky and averaging the 2D files together \citep{pope_2007}.
Spectral extraction was performed using a 2 pixel window in SPICE and data were calibrated using the same extraction window on a standard star spectrum.
For more details on the observations and data reduction for the data from GO2-20456, see \citet{pope_2007}.\\
}
\indent{
The data from PID 262 were taken in spectral mapping mode. As for the GO2-20456 data, the latent charge build up was fit and rogue pixels cleaned. A spectral cube was
generated using the CUBISM tool \citep[e.g.][]{smith_2007b}. We ensured that we did not include the slit loss 
correction factor (SLCF) while making the cube because the sources in the field of view
are point sources while the SLCF is used to correct for light diffracted into the slit due to a source which is extended wider than the spectrograph slit. The spectral
cube was generated on a 2.5$\arcsec$ spatial grid. The spectrum for each source was extracted in a 3$\times$3$\arcsec$ square aperture around the source. 
The flux density in the spectrum was then cross-calibrated with the 16 and 24 $\mu$m photometry available for the entire GOODS field.\\
}
\indent{
We know that the 3.3$\mu$m PAH signature, present in the IRAC band, will contribute only a few percent of the total IRAC photometry which is dominated by stellar photospheric emission. 
Therefore, the stellar continuum in these 21 objects needs to be accurately estimated.
We use some of the deepest data currently available in GOODS-N, including 3.6, 4.5, 5.8, 8.0 $\mu$m photometry obtained with the \textit{Spitzer} Infrared Array Camera \citep[IRAC,][]{fazio_2004}, and the F435W($B$), F606W($V$), F775W($i$), F850LP($z$) photometry obtained with the \textit{HST} Advanced Camera for Surveys (ACS) \citep[see][]{dickinson_goods}.
In order to constrain the slope of the stellar continuum in the IRAC passbands which are unaffected by the rising dust continuum, we need to apply a redshift cut corresponding to $z>0.5$.
Moreover the redshifting of the 3.3$\mu$m PAH feature out of the IRAC passbands implies a second redshift cut at $z<1.8$.
These constraints reduced the initial sample to 11 objects.\\
}
\indent{
Optical and NIR photometry must be unaffected by neighboring sources since they could contaminate the IRAC flux densities by a larger amount than the PAH signature we are trying to detect.
Using a visual analysis of optical images we rejected every source with a neighbor closer than 2 IRAC Full Width Half Max (i.e $\sim 3\arcsec$).
This reduces our sub-sample of 11 down to only 6 isolated galaxies (see Figure \ref{fig:imagette}). Of these 6, only MIPS 3419 and MIPS 5581 are from the spectral mapping observations in PID 262.
Multiwavelength photometry of these galaxies are shown in Table \ref{tab:sample}.
The optical photometry are MAG\_AUTO values from the public GOODS-N catalogs. 
The infrared photometry is measured in 4$\arcsec$ diameter apertures, with appropriate aperture corrections for the wings of the point spread function. 
Photometric uncertainties are negligibly small (except for MIPS 3419), since the galaxies are very bright, and are dominated by a 5\% systematic calibration uncertainty \citep{sirianni_2005,reach_2005}.
\\
}
\indent{
To determine the total infrared luminosity of these galaxies, we include the photometric constraints
available
from {\it Spitzer} $16\,\mu$m, $24\,\mu$m, and $70\,\mu$m observations from \citet{teplitz_2006}, Chary et al. (in prep), and \citet{frayer_2006} respectively.
In order to test the presence of an AGN, we also
consider Very Large Array (VLA) 1.4\,GHz observations (Morrison et al. in prep), 
and \textit{Chandra} X-ray observations \citep{alexander_2003}\footnote{For a detailed description of all the ancillary data existing in the GOODS fields we refer to the GOODS public webpage (http://www.stsci.edu/science/goods/)}.\\
}
\indent{
All galaxies have measured spectroscopic redshifts. 
For MIPS 4, MIPS 6, and MIPS 5581, redshifts are 
taken from the Team Keck Treasury Redshift Survey \citep[TKRS,][]{wirth_2004}, 
for MIPS 5 and MIPS 7 redshifts are from \citet{cowie_2004} while 
for MIPS 3419 we derived a spectroscopic redshift from its IRS spectra using the 9.7\,$\mu$m
silicate absorption feature. 
Five galaxies (the ones with ground-based spectroscopic redshift, i.e MIPS 4, 5, 6, 7 and 5581) are at $0.64 < z < 0.84$, while the sixth, with the IRS redshift (i.e MIPS 3419), is at $z=1.70$.
 The redshift for MIPS 3419 was measured by aligning a local galaxy template (Mrk231) whose mid-IR spectral shape is similar to the
extracted spectrum and redshifting it till it matches the observed spectrum. Since the 9.7\,$\mu$m feature is quite shallow in MIPS 3419,
the range
of plausible redshifts in agreement with the data is $1.6<z<1.9$. This is consistent with a visual analysis of the extracted spectrum which shows the flux density
dropping to zero at $\sim$25\,$\mu$m corresponding to the 9.7\,$\mu$m silicate feature at $z\sim1.6$.
The galaxies studied in this paper are star forming galaxies since their \textit{Spitzer} mid-infrared spectra display strong  $7.7\,$and$\,11.3\,\mu$m PAH emission. 4 out of 6 sources which
have adequate wavelength coverage in their mid-infrared spectrum even display strong 6.2\,$\mu$m
emission.
We fit their IRS spectra together with their broadband emission at $16$, $24$, $70\,\mu$m, and 1.4 GHz (see Table \ref{tab:radio}) with the SED library of \citet{chary_2001} \citep[see][for details]{pope_2007}.
We note that no aperture corrections are necessary since both the spectra and broadband observations integrate the entire galaxy.
The inferred bolometric luminosities are shown in Table \ref{tab:radio} and
indicate that they are luminous infrared galaxies (LIRGs), i.e with $L_{IR}(8-1000\,\mu m)>10^{11}\ \rm{L_{\odot}}$ and $SFR\ge17\ \rm{M_{\odot}yr^{-1}}$ (using the SFR-$L_{IR}$ conversion law).\\ 
}
\section{Data Analysis}
\label{sec: data analysis}
\begin{figure}
  \plottwo{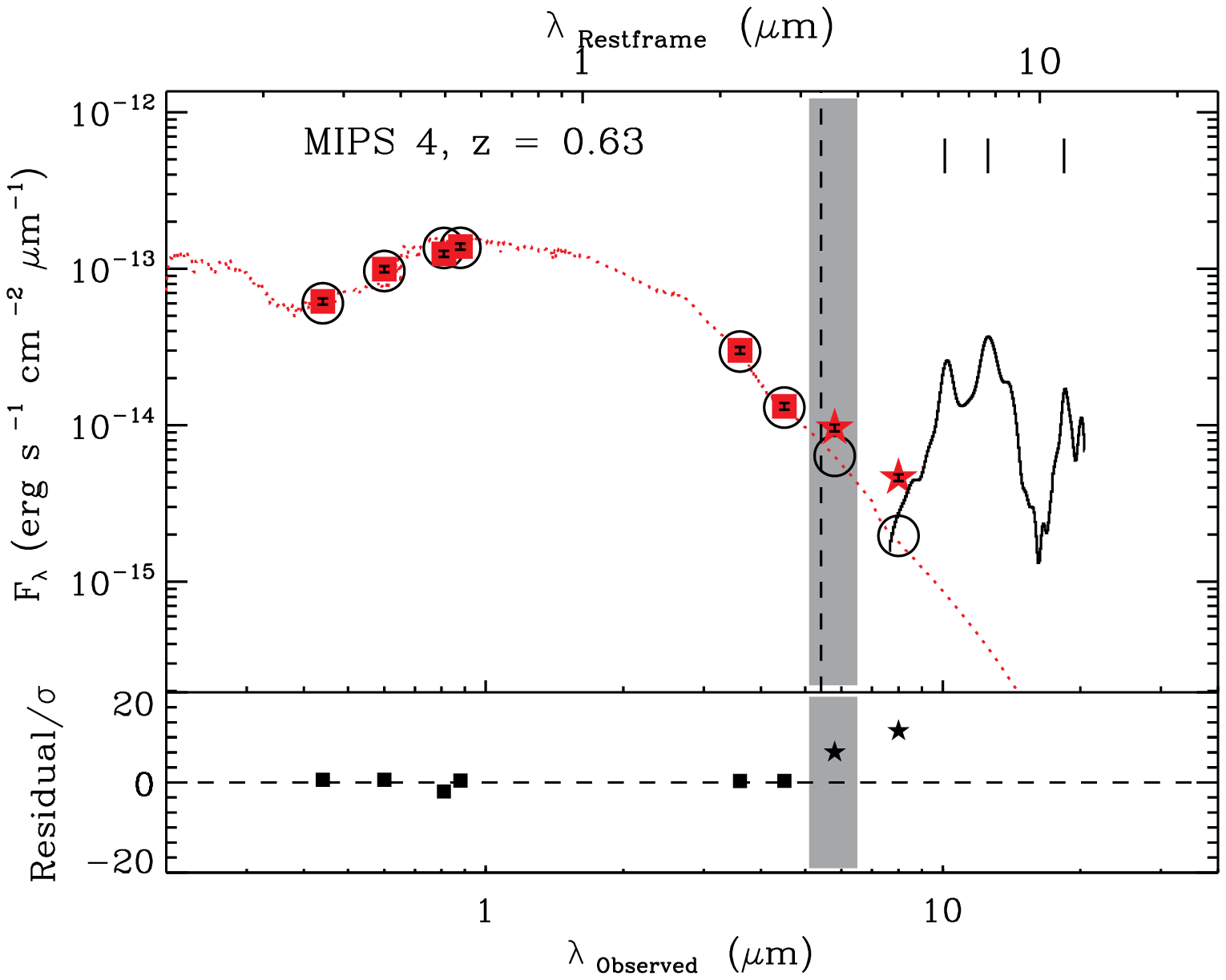}{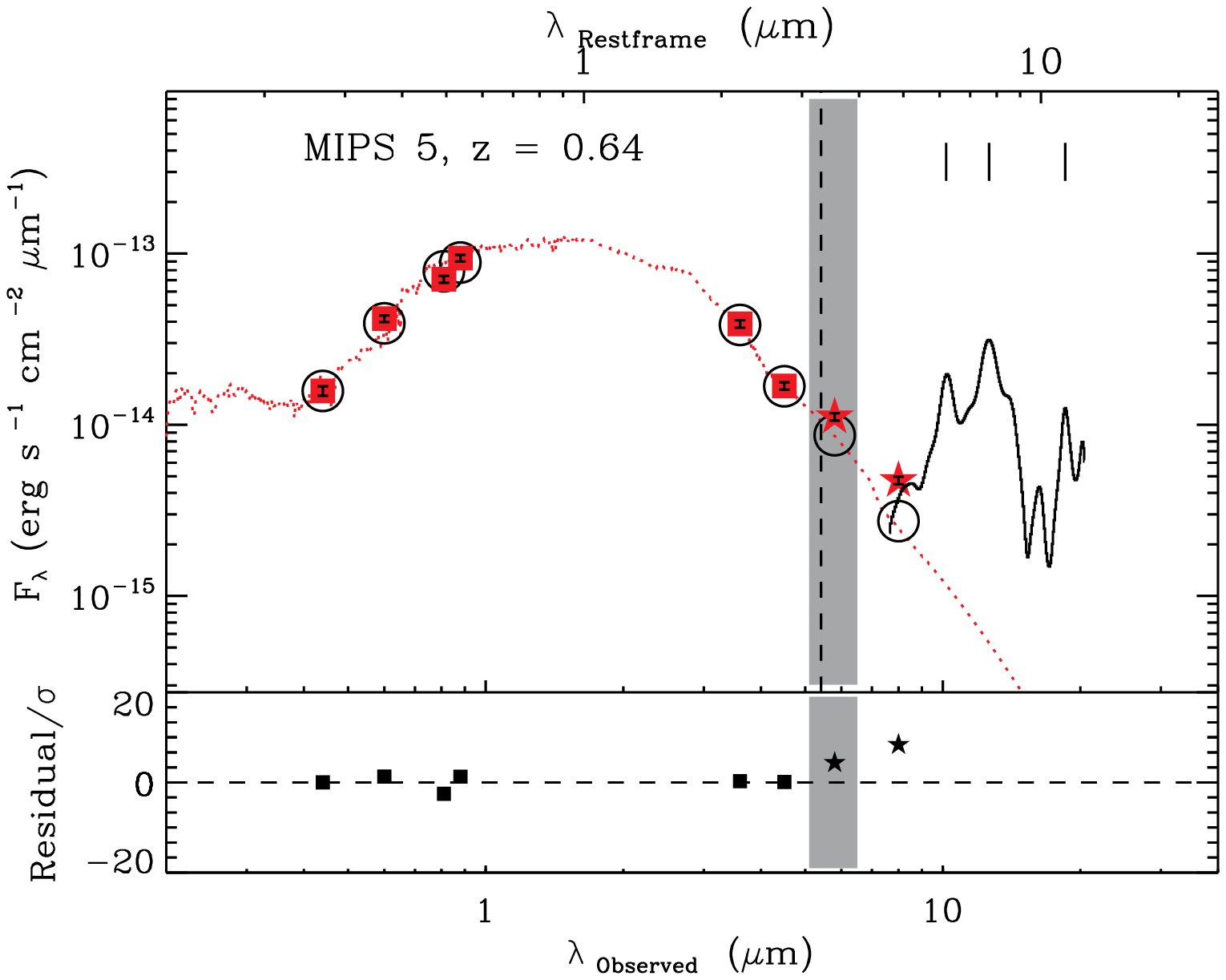}\\
  \plottwo{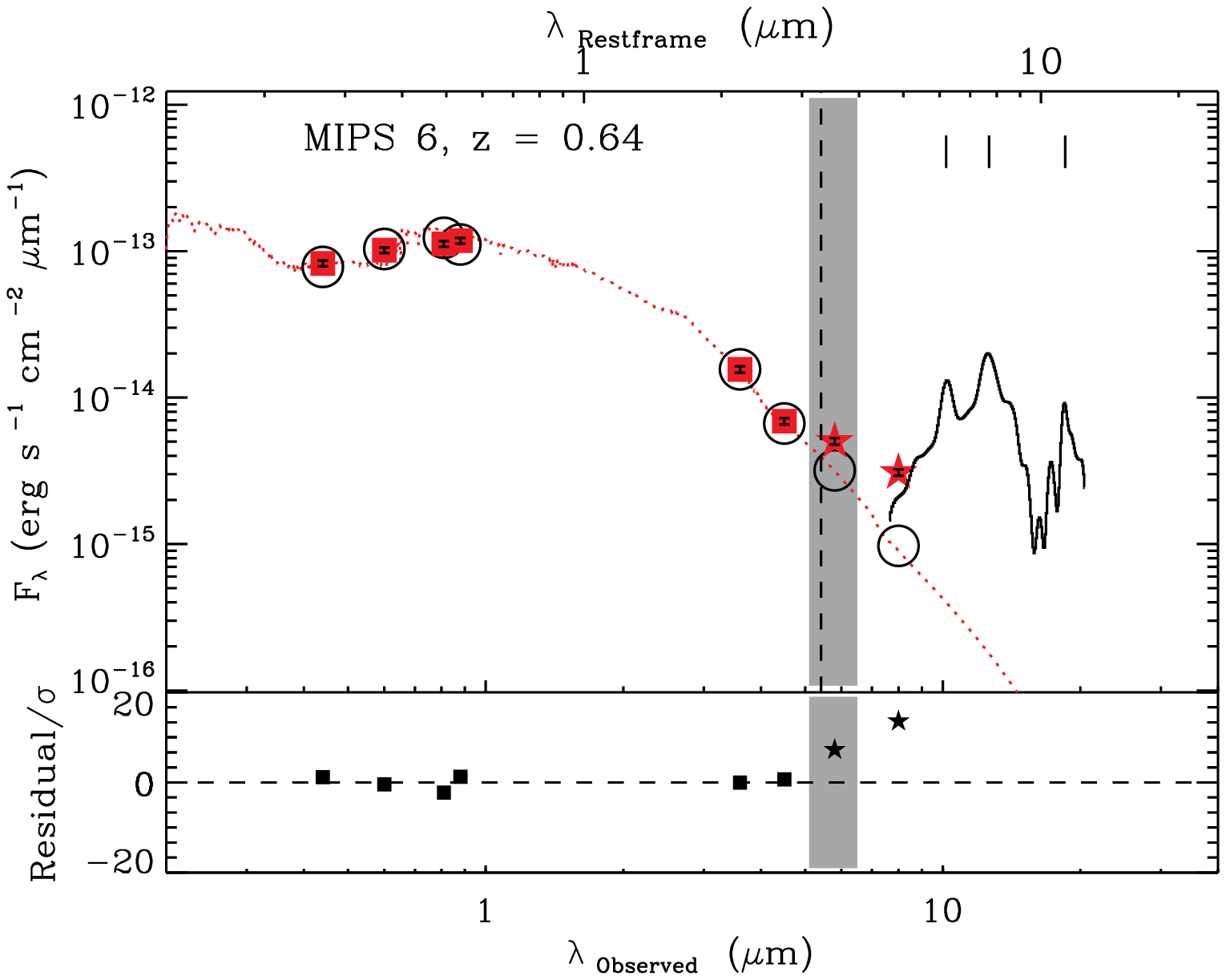}{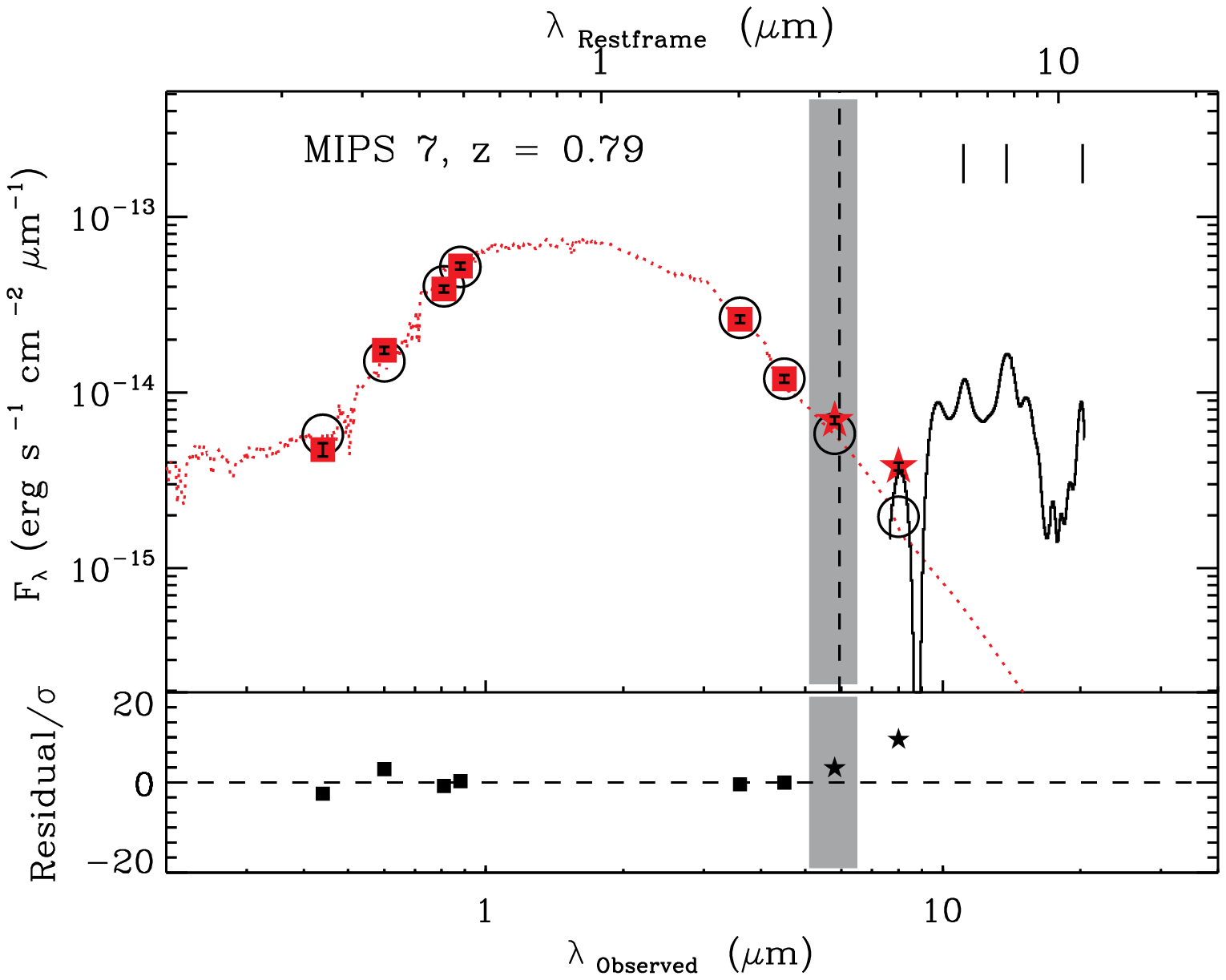}\\
  \plottwo{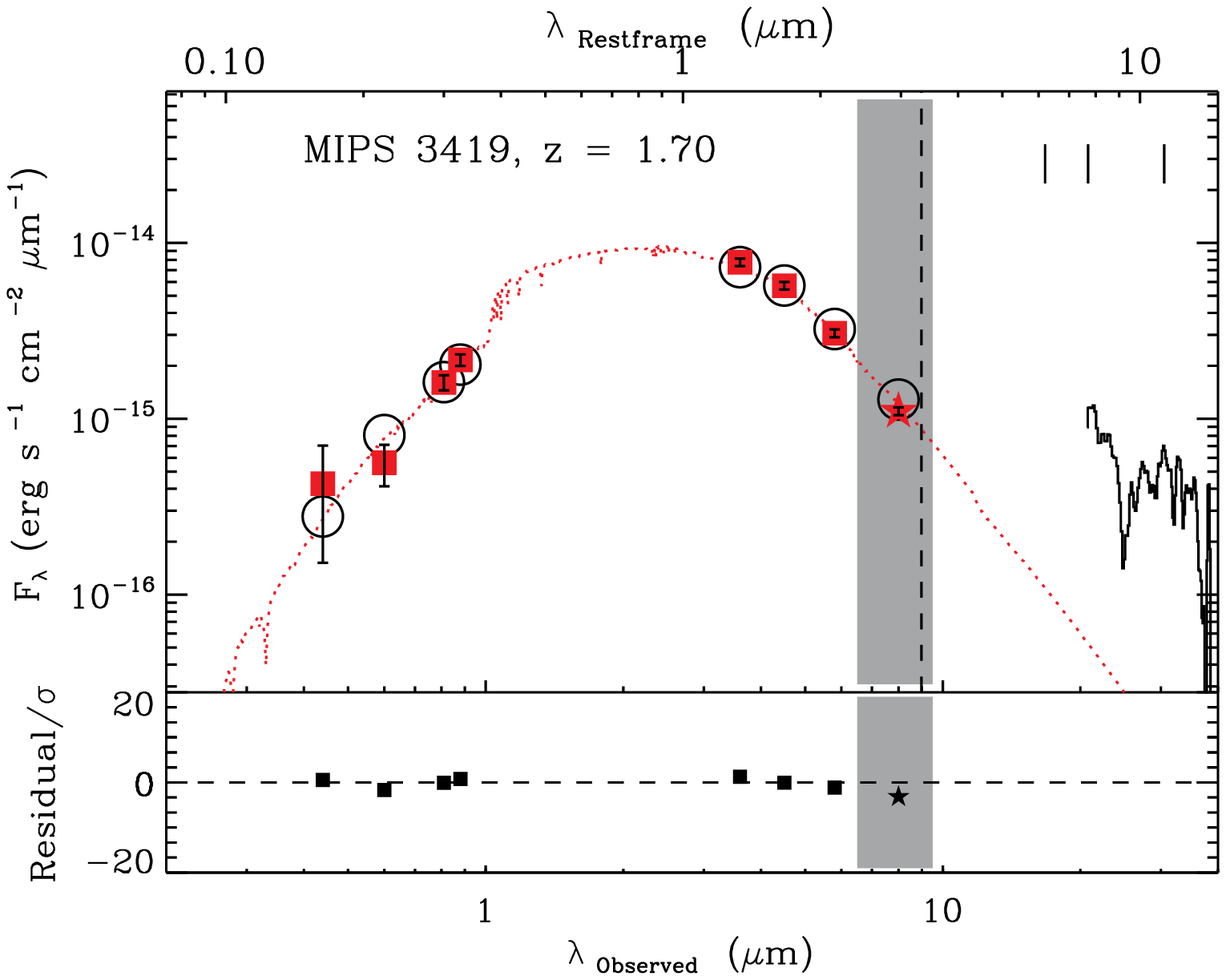}{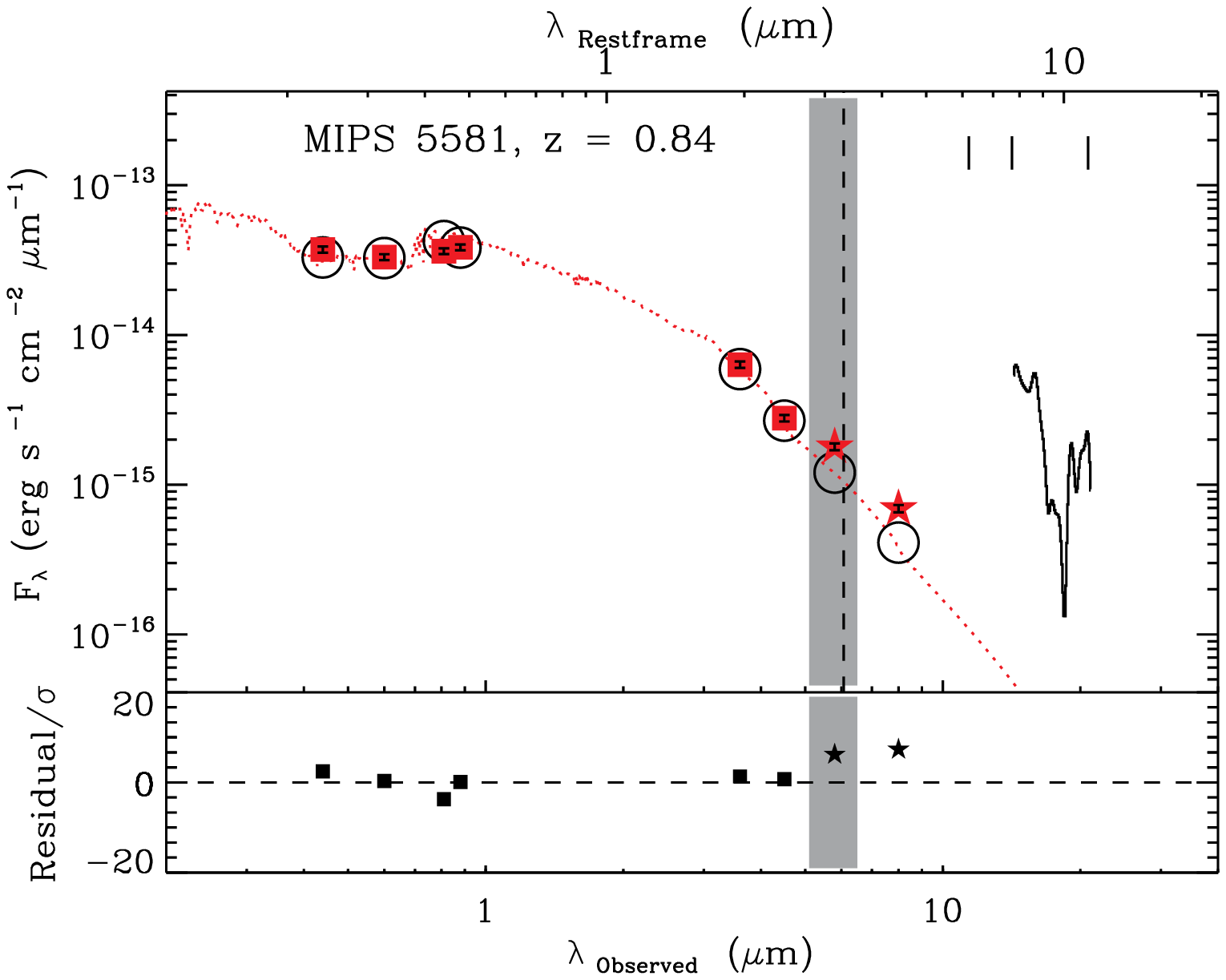}\\
  
  \caption{\label{fig:fitnormal}\footnotesize{Determination of the stellar continuum for 6 galaxies in our sample. 
	Optical and near-infrared photometry used to fit the stellar continuum are shown by red squares, near-infrared photometry excluded from the fit are shown by red stars, and the IRS spectra by a continuous line (all IRS spectra have been smoothed to the same spectral resolution of $R\sim45$ at $15\,\mu$m, using a Gaussian kernel). 
	The stellar continuum computed by PEGASE.2 is shown with the red dotted line.
The convolution of the continuum with each passband is shown as the open circle (only wavelengths blueward of the shaded region have been used to determine the fits to the stellar continuum).
The dashed vertical line represents the observed wavelength at which the redshifted $3.3\,\mu$m PAH features would be present and the shaded area represents the bandpass of the IRAC filter in which this PAH features falls.
Thin vertical lines show the location of the 6.2, 7.7, and 11.3$\,\mu$m PAH lines.
The bottom of each panel shows the ratio of the residual error, defined as the difference between the observations and the final fit, and the photometric uncertainty in each passband.
Errors obtained for the passbands used to fit the stellar continuum are shown as black squares while black stars represent errors obtained for the passbands excluded from the fit.
The difference at red wavelengths is most likely due to the 3.3\,$\mu$m PAH and the increasing contribution from the VSG continuum.
}}
\end{figure}

\indent{
We estimate the stellar continuum in the IRAC passbands by fitting stellar population synthesis models to the multi-band photometry.
The library of stellar emission used as templates is computed with PEGASE.2 \citep{fioc_1997}.
Using the range of star formation, infall, and wind histories described in Table \ref{tab:peg para},  and available on the Le PHARE website\footnote{http://www.oamp.fr/arnouts/LE\_PHARE.html} (Arnouts et al., in prep), we construct an atlas of templates spanning galaxies of all Hubble types.\\ \\
}
\indent{
For each object, original PEGASE.2 templates are shifted to the spectroscopic redshift of the source and convolved with the transmission of each observed filter.
We apply an age constraint to ensure that galaxies do not correspond to a template with an age greater than
the age of the Universe at its redshift. 
Finally we include extinction as a free parameter following a Calzetti law \citep{calzetti_1994}.\\
}
\indent{
The templates are fit to all the \textit{BViz} photometry and some but not all of the IRAC passbands.
The fit of the stellar continuum excludes the IRAC bands where the $3.3\,\mu$m PAH signature may be present and also excludes longer wavelengths which are influenced by the rising dust continuum.
Hence, for galaxies situated at $0.5<z<1$, the IRAC 5.8 and 8.0 $\mu$m photometry is excluded from the fit (i.e MIPS 4, MIPS 5, MIPS 6, MIPS 7, MIPS
5581) and for the galaxy situated at $z=1.7$ (i.e MIPS 3419),  only the IRAC 8.0 $\mu$m photometry is excluded from the fit.\\
}
\indent{
Applying all these constraints, we calculate the best fit using a $\chi^{2}$ minimization technique.
For each object the best fit yields the stellar continuum corresponding to a specific combination of 4 parameters: type of galaxy (E, Sa, Sb\dots), age, normalization (i.e stellar mass), and extinction (see Table \ref{tab:fit peg}).
The result of the individual fitting is presented in Figure \ref{fig:fitnormal}.
For all objects the residual difference between the observed photometry and the best fit is below 15\%.\\ \\
}
\indent{
Once the best fit is obtained a residual IRAC flux can be derived by computing the difference between the IRAC photometry and the stellar continuum (see Table \ref{tab:pah line}).
The resulting IRAC excess is dependent on the stellar continuum used to fit the data and thus relies on the specific value of our 4 parameters (Type of galaxy, Ages, Mass, Extinction).
To assess this dependence and to estimate the error associated to each IRAC excess that we measure, we have undertaken a Monte-Carlo approach.
For each source,
we randomly vary the photometry in each observed band by up to $\pm$3$\,\sigma_{band}$ where
$\sigma_{band}$ is the photometric uncertainty in that band.
Hence these new values for the photometry are still consistent with the observations.
We recompute the best fit SED to these new data values and re-derive the inferred IRAC excess.
For each object, this procedure is repeated 100 times.
The error is then defined as the standard deviation of these 100 values.
The result of the error estimation is presented in Table \ref{tab:pah line}.
These error bars account for both the photometric error bars and the range of stellar continuum models which fit the observed data.\\
}
\indent{
We note that the Maraston (2005) models for stellar emission cannot be used for this analysis because
the empirical spectra of thermally pulsing AGB stars, which are a key feature of the models,
do not extend beyond rest-frame 2.5\,$\mu$m.
Usage of these models would result in an inferred line flux systematically higher by a factor $\sim$2 than the one derived with PEGASE.2, due to this break at 2.5\,$\mu$m.\\
}
\indent{
The results, shown in Table \ref{tab:pah line}, indicate that for all but one galaxy (MIPS 3419), we have detected a significant excess in the IRAC $5.8\,\mu$m passband.
We note that the absence of an IRAC excess for MIPS 3419 could be due to a wrong redshift determination.
Indeed as discussed in Section \ref{sec: samples} the redshift of this source was only derived from its IRS spectrum.
In the next section, we discuss the physical origin of the IRAC excess for the remaining five galaxies.\\
}
\section{Discussion on the origin of the IRAC excess\label{sec:Origin}}
\indent{
The IRAC excess can originate from four different components: (i) hot dust emission from an obscured AGN, (ii) free-free and hydrogen recombination line emission from ionized gas, (iii) a possible continuum observed in the diffuse medium of our Galaxy \citep{flagey_2006,lu_2004} and in some local star forming galaxies \citep{lu_2003} and, finally to (iv) the $3.3 \mu$m PAH feature itself.
In the following we discuss the contribution of each of these components to the measured IRAC excess.
}
\subsection{Obscured AGN\label{sec:AGN}}
\begin{figure}
	\plottwo{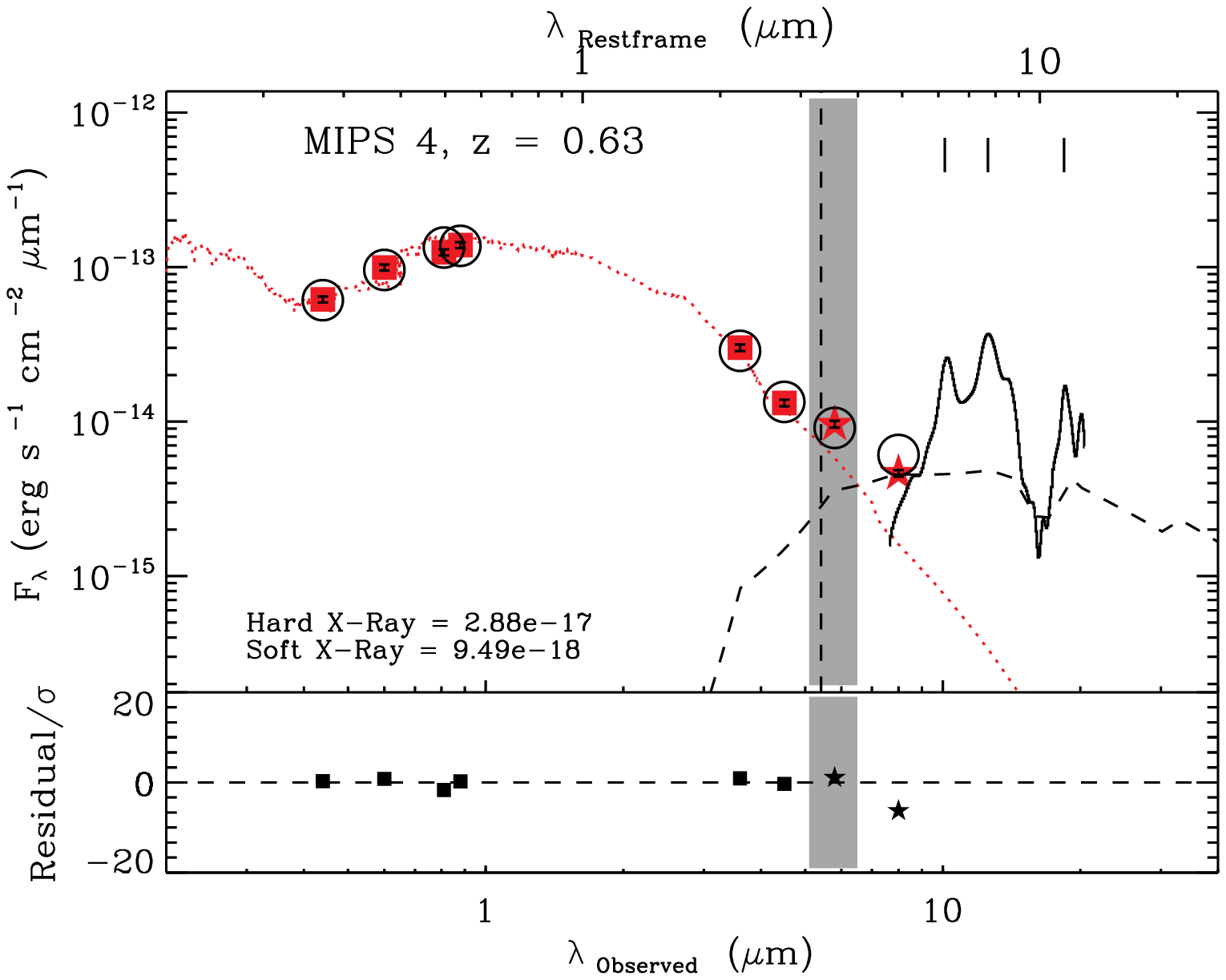}{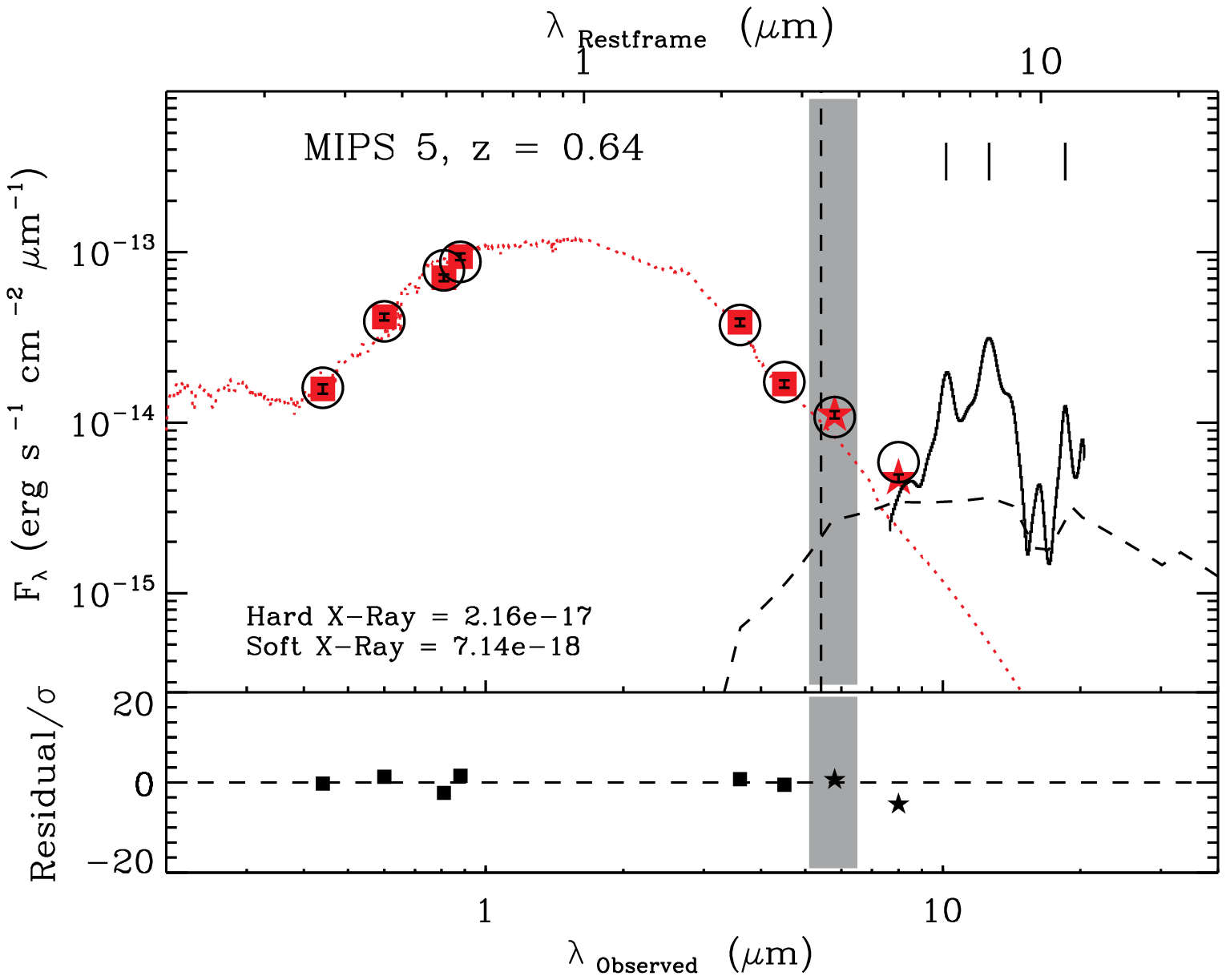}
	\plottwo{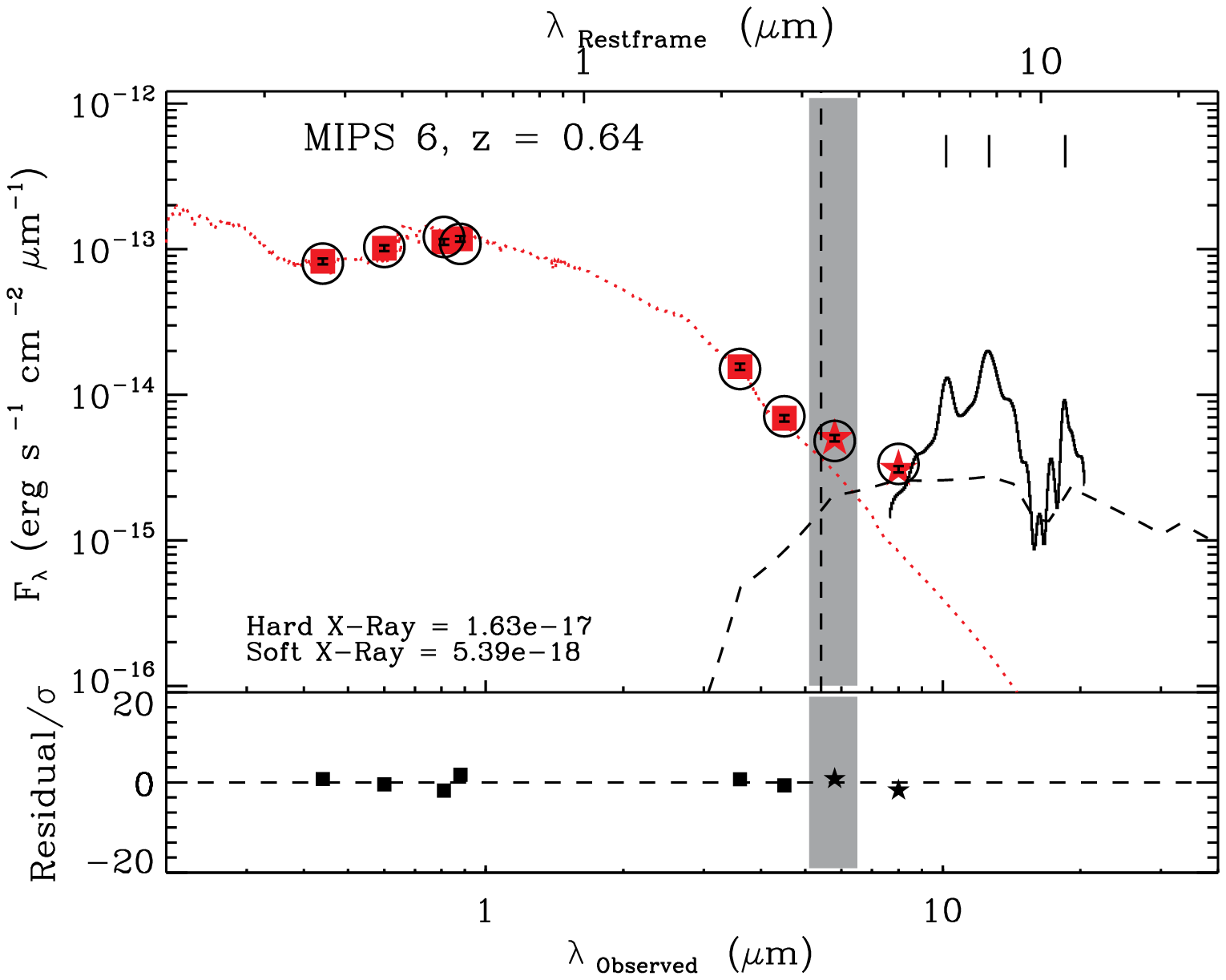}{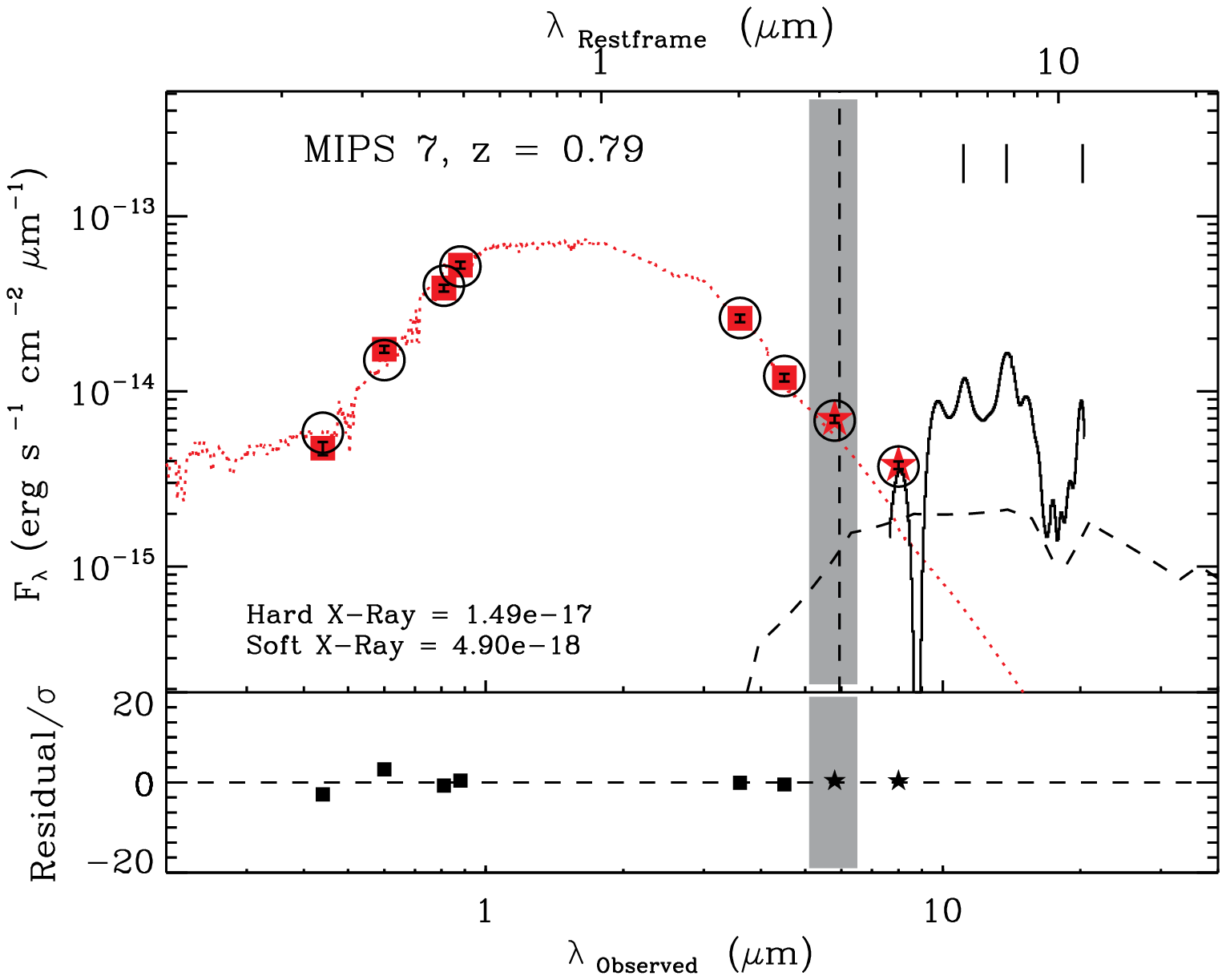}
	\plotfiddle{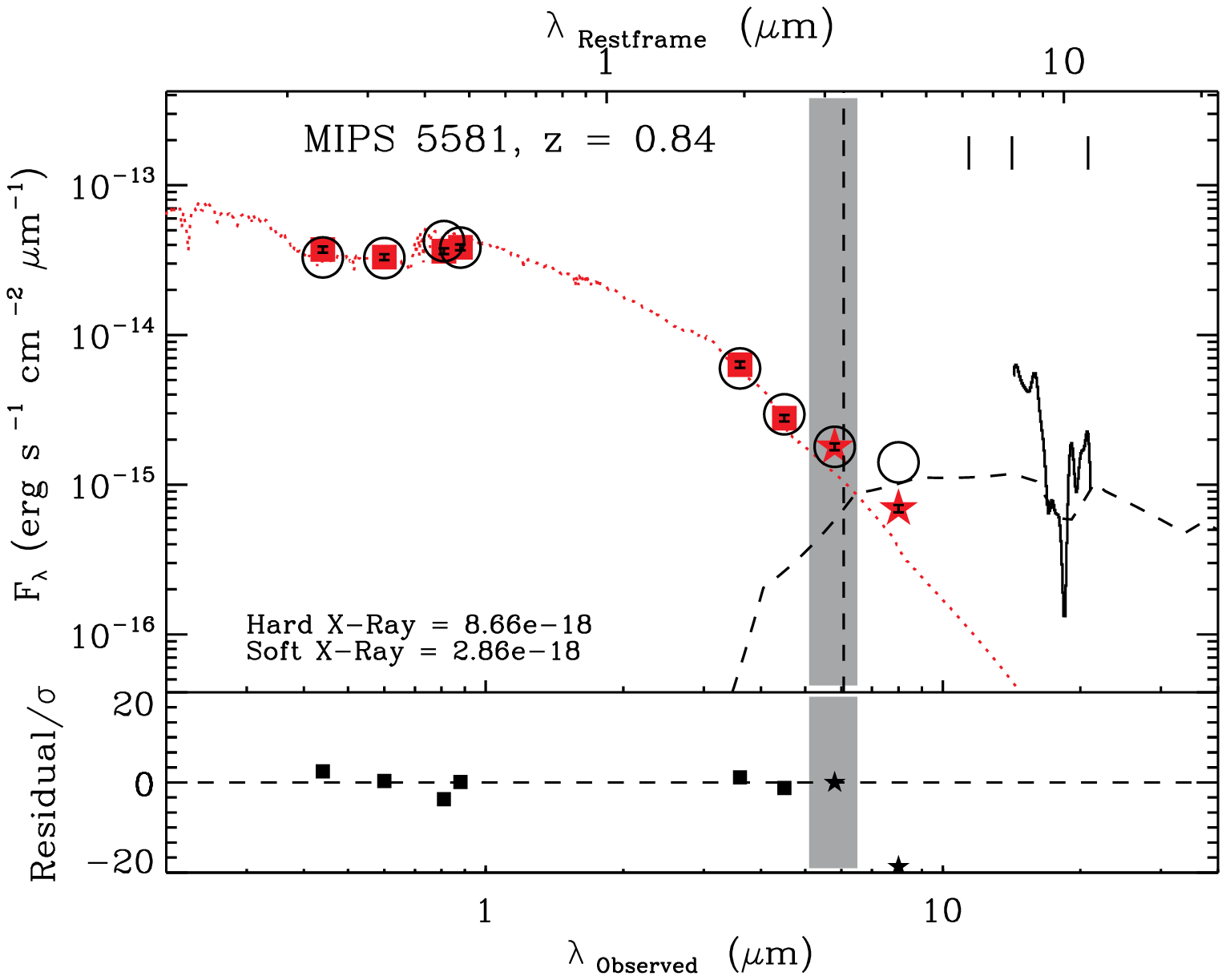}{0.pt}{0.}{210}{160}{125}{0}
	\caption{\label{fig:fitagn} \footnotesize{Determination of the stellar continuum for 5 galaxies from which we have removed an AGN contribution proportional to the IRAC excess calculated in section \ref{sec: data analysis}. Lines and symbols are the same as in Figure \ref{fig:fitnormal}.
	  The stellar continuum computed by PEGASE.2 and the AGN contribution are shown as dot and dash lines respectively (as before, only wavelengths blueward of the shaded region have been used to determine the stellar continuum).
	  The final fit, sum of the stellar and the AGN contributions, convolved through each band pass is shown as open circle.
           The X-ray contribution of the normalized AGN is shown in each panel in $\rm{erg\,s^{-1}\,cm^{-2}}$. For sources 4, 5
and 5581, the IRAC excess is inconsistent with being due to an obscured AGN (see text for details).
}}
\end{figure}

\indent{
A possible origin for the IRAC excess found in 5 galaxies of our sample could be hot dust emission from an obscured AGN.
We assess this possibility by studying the X-ray, radio, and optical properties of these galaxies.\\
}
\indent{
Using \textit{Chandra} observations in GOODS-N \citep{alexander_2003} we find that three galaxies of our sample (MIPS 4, 6, and 7) are detected in the soft X-rays (0.5-2 \rm{KeV}) and only one in the hard band (2-8 KeV) (MIPS 6; see Table \ref{tab:X-ray}).
The photon index $\Gamma= 1.06$ of MIPS 6 indicates the presence of an obscured AGN.
Moreover, the fact that the
resolved fraction of the X-ray background \citep[CXB,][]{worsley_2004} decreases with increasing energy 
suggests the existence of a population of obscured AGN which might be
missed in even the deepest X-ray surveys \citep{barger_2007}.
To address the presence of such AGNs we compare the SED of the galaxies in our sample with the SED of the Compton thick AGN in NGC 1068.
The choice of NGC 1068 has been made since its SED follows a $F_{\lambda} \propto \lambda^{-\alpha}$ power-law with $\alpha=2.25$ in the NIR. This is a typical spectral index 
for obscured AGN \citep{risaliti_2006}.\\
}
\indent{
The SED for the nucleus of NGC1068 was derived from the high spatial resolution ground-based
photometry compiled by \citet{Galliano_2003},
the ISOPHOT-S spectrum presented by \citet{Rigopoulou_1999}, and the IRAS broadband photometry.
The IRAS
large beam photometry at 12, 25, 60, and 100\,$\mu$m was fit with a far-infrared curve
comprising of multiple temperature dust components as in \citet{chary_2001}.
This synthetic spectrum was normalized at 12\,$\mu$m to the
ISOPHOT-S spectrum of the nucleus which extends between 6$-$12\,$\mu$m \citep{Rigopoulou_1999}.
The ISOPHOT-S spectrum is clearly dominated by continuum emission from the AGN. 
However, the flux density appears to systematically exceed the high resolution ground-based
photometry by a factor of 2. It is unlikely that this is due to extended emission in the vicinity
of the nucleus which enters the ISOPHOT-S aperture
since the spectrum is clearly dominated by hot dust emission. It is more
likely due to flux calibration uncertainties associated with the ISOPHOT-S spectrum.
After dividing the integrated ISOPHOT+IRAS spectrum by a factor of 2, bringing all the data in 
agreement, we extend the spectrum, using
the ground-based photometry, down to 1\,$\mu$m.\\ 
}
\indent{
Assuming that the excess observed in the IRAC passband is only due to hot dust from an obscured AGN, we calculate the normalization factor for the NGC 1068 SED corresponding to this excess.
Then subtracting the AGN contribution from the multi-wavelength photometry and using the fitting procedure described in section \ref{sec: data analysis}, we fit the revised photometry with the PEGASE.2 population synthesis model.
The result of these fits are presented in Figure \ref{fig:fitagn}.\\
 }
\indent{
For 3 objects, MIPS 4, 5, and 5581, these fits yield a predicted flux in the IRAC $8\,\mu$m passband which exceeds the observed photometry by a factor 6.2, 4.7, and 18.6 $\sigma$ respectively.
These over-estimations of the IRAC $8\,\mu$m photometry prove that MIPS 4, 5 and 5581 cannot harbor such an obscured AGN since their IRAC color (IRAC5.8/IRAC8.0) do not follow 
the typical colors of hot dust emission in NGC 1068.
Moreover to match the IRAC $8\,\mu$m photometry the obscured AGN, if present in these 
galaxies, would have to be at least a factor of 1.3, 1.25, and 2.0 less luminous 
respectively.
Then these obscured AGN would not be luminous enough to explain the $5.8\,\mu$m IRAC excess.
On the contrary, for MIPS 6 and 7, the renormalized NGC 1068 SED results in an IRAC 8 $\mu$m flux which is consistent with observations and thus suggests that hot dust from an obscured AGN could explain the IRAC excess.
One should note that these conclusions are dependent on the NIR spectral index of NGC 1068 (i.e $\alpha=2.25$).
To assess this dependence we also attempted to fit the IRAC excess using the SED of Mrk 231 \citep{Armus_2007} which has a low spectral index of $\alpha=0.25$.
We find that the AGN fit exceeds the observed $8\,\mu$m IRAC flux for MIPS 4, 5, and 5581 by a value of 4.1, 3.0, and 9.9 $\sigma$ respectively.
Although these values are a factor of two smaller than the ones derived for NGC 1068 they also weaken the possibility that hot dust emission can explain the IRAC excess.\\ \\
}
\indent{
We now evaluate if the observed X-ray emission from these sources is consistent with this simple model.
We assume that the X-ray emission is the sum of starburst and AGN contributions, and compare with the \textit{Chandra} observations (see Table \ref{tab:X-ray}).
Since our galaxies show strong PAH features, the bolometric luminosity calculated in Section 2, is 
dominated by  star-formation.
We calculate the starburst X-ray contribution using the relationship between SFR and $L_{IR}$ \citep{kennicutt_1998} and the relationship between SFR and $L_{0.5-8keV}$ \citep{bauer_2002}:
}
\begin{equation}\label{eq:sfr lir}  
	SFR[\rm{M_{\odot}\,yr^{-1}}]=4.5\times10^{-44}L_{IR}[\rm{erg\,s^{-1}}]
\end{equation}
\begin{equation}
	SFR[\rm{M_{\odot}\,yr^{-1}}]=1.7\times10^{-43}L_{0.5-8\rm{keV}}^{1.07}[\rm{erg\,s^{-1}}]
\end{equation}
\indent{
Using $\Gamma=1.9$ as the typical starburst photon index we estimate the soft and hard X-ray rest-frame luminosity ($L_{soft/hard}^{restframe}$) predicted for the starburst. The observed flux in each band ($f_{soft/hard}^{observed}$) is then derived by applying a \textit{K}-correction give by Equation (2) of \citet{bauer_2002}:
}
\begin{equation}  
	\label{eq:kcorrec}
	f_{soft/hard}^{observed}[\rm{ erg\,s^{-1}\,cm^{-2}}]=\frac{L_{\it{soft/hard}}^{\it{restframe}}[\rm{erg\,s^{-1}}]}{4\pi d_{l}^{2}\,(1+z)^{\Gamma -2}}
\end{equation}
\indent{
where $d_{l}$ is the luminosity distance. 
For the AGN contribution, we adopt $L_{2-10 \rm{keV}}=2.8\times 10^{-12} \rm{erg\,cm^{-2}\,s^{-1}}$ and $\Gamma=1.2$ \citep{ogle_2003} for NGC 1068.
From this, we predict the emission of our objects in the soft and hard X-ray bands (cf Figure \ref{fig:fitagn}) using the NGC 1068 normalization factor calculated from the IRAC excess and the \textit{K}-correction given by Eq. \ref{eq:kcorrec}.\\ \\
}
\indent{ 
For all galaxies the total (AGN + Starburst contribution) soft X-ray emission, dominated by the starburst contribution ($L_{soft}^{Starburst}/L_{soft}^{AGN}\thicksim10$), is in agreement, within the error, with the observed soft X-ray flux by a factor of 1.3, 1.7, 1.8, and 1.0 for MIPS 4, 5, 6, and 7 respectively and under the \textit{Chandra} threshold for MIPS 5581 (see Table \ref{tab:X-ray}).\\
}
\indent{
For MIPS 4, 5, 7, and 5581, the total hard X-ray prediction is under the \textit{Chandra} threshold and thus in agreement with observations.
For MIPS 6, our hard X-ray prediction ($1.48\times 10^{-16}\,\rm{erg\,s^{-1}\,cm^{-2}}$) is a factor of 1.7 below the observation ($2.54\times 10^{-16}\,\rm{erg\,s^{-1}\,cm^{-2}}$) but still, within the error, in agreement.
This suggests that the obscured AGN harbored by this galaxy is significantly more luminous than for NGC 1068 in the X-rays.\\
}
\indent{
In summary, the X-ray analysis reveals the presence of an obscured AGN in MIPS 6 but is not conclusive for MIPS 4, 5, 7, and 5581 since either the presence or absence of an obscured AGN would be in agreement with the data.
The X-ray predictions using the Mrk 231 SED are similar to those derived using the SED of NGC 1068.
Hence, our conclusions are independent of the AGN SED adopted to fit the data.\\ \\
}
\indent{
Using the Chary \& Elbaz library which follows the local radio/far-infrared correlation we predict $L_{1.4\rm{GHz}}$ using the $24\,\mu$m luminosity (see Table \ref{tab:radio}).
For MIPS 4, 5, and, 5581 we find $L_{1.4GHz}^{predicted}/L_{1.4\rm{GHz}}^{obs} \thicksim 0.8$, and for MIPS 6 and MIPS 7, which may be powered by an AGN,  we find $L_{1.4\rm{GHz}}^{predicted}/L_{1.4GHz}^{obs} \thicksim 1.2$ and $0.8$ respectively.
This agreement between predictions and observations suggests that the infrared emission of all these galaxies is dominated by star-formation.
Finally, assuming a spectral index ($\alpha$ in F$_{\nu}\propto\nu^{-\alpha}$) of 0.8 we calculate for each galaxy $L_{1.4\,\rm{GHz}}^{restframe}$ (see Table \ref{tab:radio}).
These values are then compared to $5\times10^{23} \rm{W.Hz^{-1}}$ which is the typical radio luminosity used to distinguish between a starburst and an AGN population at $z \thicksim 0.7$ \citep{yun_2001,cowie_2004}.
For each galaxy in our sample, $L_{1.4\,\rm{GHz}}^{restframe}$ is below $5\times10^{23} \rm{W.Hz^{-1}}$ which is consistent with the assumption that our galaxies are star-formation dominated systems.\\ \\
}
\indent{
Finally, we analyze the optical spectra of three galaxies (MIPS 4, 6, and 5581) available in the TKRS database \citep{wirth_2004}.
No clear evidence of high ionization lines like [Ne III] or [Ne V] or large [OIII]/$H_{\beta}$ ratio are found arguing against the possibility that the IRAC excess of these three galaxies is dominated from AGN activity.
However, we note that this analysis includes MIPS 6 which shows clear evidence of an AGN in its X-ray emission.
\\ \\
}
\indent{
The conclusion of the SED fitting, X-ray, optical and radio diagnostics on the presence of an obscured AGN in these galaxies can be summarized as follows:\\
} 
\indent{
- For MIPS 4, 5, and, 5581 the disagreement between the IRAC $8\,\mu$m expected for a prototypical obscured AGN and the observations suggests that these galaxies do not harbor an AGN. Moreover for MIPS 4 and 5581, this conclusion is also supported by the agreement found between their radio and mid-infrared luminosities as well as the absence of AGN signature in their optical spectra.\\
}
\indent{
- For MIPS 7, the IRAC excess could be reproduced by a buried AGN but we found no other evidence for an AGN neither from its optical nor its X-ray properties.\\
}
\indent{
- Finally, MIPS 6 presents marginal evidence for the presence of an obscured AGN in its X-ray emission (1.7 times larger than typically expected for star-formation).
Its radio luminosity and optical spectrum are consistent with a star forming dominated system.
In the following, we will discuss the case of this galaxy separately since we cannot rule out its contamination by an obscured AGN.
}
\subsection{Free-free and gas lines\label{sec:freefree}}
\indent{
The total contribution of free-free and recombination
line emission have been estimated by \citet{flagey_2006} to be about 1\% and 3\% in the IRAC 3.6 $\mu$m (i.e the residual IRAC 5.8 $\mu$m emission for MIPS 4, 5, 6, 7 and 5581 or the residual IRAC 8.0 $\mu$m for MIPS 3419) and in the IRAC 4.8 $\mu$m (i.e the residual IRAC 8.0 $\mu$m for MIPS 4, 5, 6, 7 and, 5581) channels respectively.
Even in the extreme case where these contributions would reach 3\% and 11\% \citep{flagey_2006}, the free-free 
and line emission contribution to the IRAC excess would be negligible.
Furthermore if the IRAC excess was due to free-free emission, it would be accompanied by a comparably high value at shorter
wavelengths (i.e in the 4.5 $\mu$m IRAC channel) due to the blue spectrum of free-free emission. 
The flux density
in the 3.6 $\mu$m and 4.5 $\mu$m IRAC channels in MIPS 4 ,5, 6, 7 and 5581 are consistent with being dominated by stellar emission (see Figure \ref{fig:fitnormal}) excluding
this possibility.
}
\subsection{Warm dust continuum\label{sec:cont}}
\indent{
A continuum underlying the $3.3 \mu$m PAH feature has been detected in the Galactic diffuse medium \citep{flagey_2006,lu_2004} and in some local star forming galaxies \citep{lu_2003}. 
There, the NIR continuum, well fitted by a modified black body with temperature spanning the range $700-1500$ K and a $\lambda^{-2}$ emissivity law, contributes to about 70\% of the IRAC 3.6 $\mu$m flux.
They also conclude that the IRAC 4.5 $\mu$m channel is totally dominated by this NIR continuum.\\
}
\indent{
Assuming that the IRAC excess, that we measure here in distant galaxies, is due to such a continuum, the $f_{\nu}^{predicted}(5.8\,\mu m)/f_{\nu}^{predicted}(8.0\,\mu m)$ color for MIPS 4, 5, 6, 7 and, 5581 would span the range [0.97 - 2.07].
We can assume these color predictions to be lower limits since they assume no contribution of the $3.3 \mu$m PAH feature in the IRAC 5.8 $\mu$m passband.
The observed ratios after subtracting the stellar contribution (i.e $[f_{\nu}^{observed}(5.8\,\mu m)-f_{\nu}^{PEGASE.2}(5.8\,\mu m)]/[f_{\nu}^{observed}(8.0\,\mu m)-f_{\nu}^{PEGASE.2}(8.0\,\mu m)]$) are 0.64, 0.64, 0.45, 0.32 and 1.08 for MIPS 4, 5, 6, 7 and 5581 respectively.
Although, one galaxy is marginally consistent with this warm dust continuum (MIPS 5581), the remaining four correspond to a modified black body temperature of 550 K, hence colder than the temperature range of the NIR continuum found in \citet{flagey_2006}.
As a result it is unlikely, unless the shape of the continuum were different from anything known, that the IRAC 5.8 $\mu$m passband is dominated by the same NIR continuum in MIPS 4, 5, 6 and, 7.\\
}
\indent{
The physical origin of such a high temperature continuum remains unclear and we cannot totally rule out that it reaches a different temperature range in distant highly star forming galaxies. 
Near-infrared spectroscopy (L-band, 3Ð4\,$\mu$m) of a sample of 24 local actively star forming galaxies \citep{imanishi_2000} does exhibit some contribution from a hot continuum which could, in principle, serves as a better reference for the present sample of distant star forming galaxies.
We find that their continuum can be reproduced by a colder modified black-body than the one found in \citet{flagey_2006} since it reaches a temperature of T$\sim$ 400 K.
This colder continuum, which is more consistent with the IRAC temperature of the present sample, would contribute for 60\% of the IRAC excess.
However, this sample consists only of the nuclei of nearby ultraluminous infrared galaxies (ULIRGs) obtained using the Subaru IRCS near-infrared spectrograph.
These galaxies were chosen as candidate AGNs, the goal of the authors being to look for buried AGNs, which they claim to find with various intensities in all objects.
Hence the warm dust continuum observed in their sample can be highly contaminated by AGN activity.\\
}
\subsection{$3.3 \,\mu$m PAH feature\label{sec:PAH}}
\begin{figure}
	\plotone{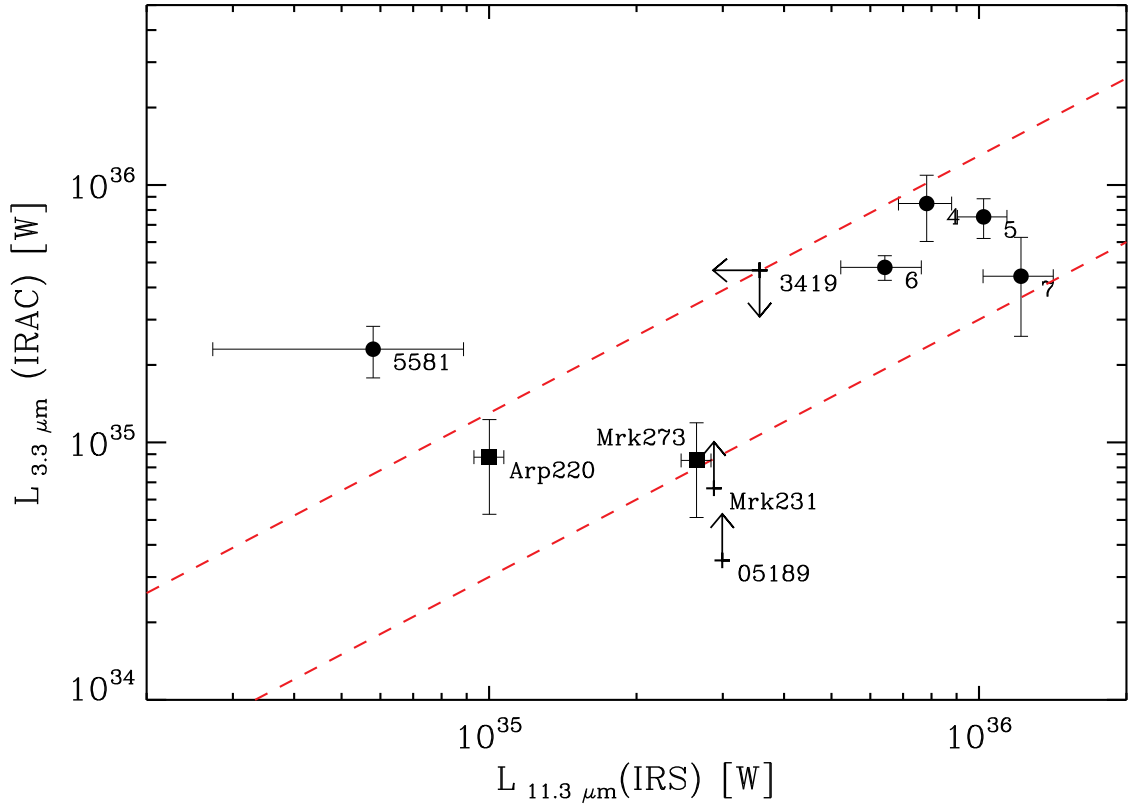}
	\caption{\label{fig:line correlation} 
	\footnotesize{Comparison between the inferred $3.3\,\mu$m PAH line flux and the $11.3\mu$m PAH line flux measured from the IRS spectra. 
	Uncertainties for $L_{3.3\,\mu m}$ are computed using a Monte Carlo approach (see text for details). The arrow corresponds to a $2\sigma$ limit. 
	The dashed lines represent $L_{3.3\,\mu m}=\alpha\, L_{11.3\mu m}$ with $\alpha = 0.3$ or $1.3$ for the ionized and neutral PAH respectively \citep{li_2001}.
	For comparison, we plot the PAH luminosities of 2 local starburst galaxies (Arp 220 and Mrk 273) and 2 local AGN (Mrk 231, IRAS 05189-2425) with $L_{3.3\,\mu m}$ from \citet{imanishi_2006} and $L_{11.3\mu m}$ from \citet{Armus_2007}.}}		
\end{figure}

\indent{
Assuming that the IRAC excess is due to the 3.3\,$\mu$m PAH feature, we show that the inferred line fluxes are consistent with expectations from dust models.
To infer the $3.3 \,\mu$m PAH line flux from the IRAC excess, we adopt a Gaussian profile for the PAH line.
Then we convolve this profile with the IRAC filter curve to obtain the normalization factor for the Gaussian profile corresponding to the IRAC excess.
Finally the true flux of the $3.3 \mu$m PAH line is obtained by integrating over the normalized Gaussian profile.
The estimated value of the $3.3\,\mu$m PAH line flux calculated for each galaxy is shown in Table \ref{tab:pah line}.
We note that the line flux inferred by adopting a Drude profile is only 1.12 times the value derived for a Gaussian profile.
Therefore, our conclusion does not depend on this choice.
We also note that the extinction derived from the stellar population
fits to the photometry (see Table \ref{tab:fit peg}) reaches an average value of $A(3.3\ \mu m)\approx 0.14$.
Hence, the extinction correction to the inferred $3.3\,\mu$m PAH lines fluxes is smaller than the
uncertainty in the flux and are therefore negligible.\\
}
\indent{
Models predict that the $3.3\,\mu$m and the $11.3\,\mu$m lines originate in the same C-H bond \citep{li_2001, duley_1981}  which results in a linear correlation between these two lines.
Hence, finding a correlation between the inferred $3.3\,\mu$m line flux and $11.3\,\mu$m line flux, measured from the IRS spectra, would support the assumption of a negligible continuum.\\
}
\indent{
Using the \citet{draine_2007} templates and a Gaussian profile for the PAH features, we find $L_{3.3\,\mu m}=\alpha\, L_{11.3\mu m}$ with $\alpha = 0.4\pm 0.2 $ in the models.
However these templates do not represent the extreme case where the PAH are totally ionized or neutral.
Using Table 1 of \citet{li_2001}, we find $\alpha = 0.3$ or $1.3$ for the ionized and neutral PAH respectively.\\
}
\indent{
The $11.3\mu$m line flux of each galaxy has been measured in its IRS spectrum using the ISAP \citep{higdon_2004} component of SMART (see Table \ref{tab:pah line}).
The line fit is done assuming a Gaussian profile for the PAH line and a constant continuum across the line.
The uncertainty in the line flux is dominated by the signal-to-noise and wavelength
coverage in the IRS spectrum which makes the derivation of the continuum slope difficult.
In the case of MIPS 7, where the wavelength coverage does not encompass the entire 11.3$\mu$m PAH line, we simulate the line flux uncertainty by using spectra of low redshift galaxies \citep{Armus_2007} and applying a similar wavelength cutoff.
We find that the maximum uncertainty in the derivation of the line flux is due to the unknown shape of the underlying continuum and is at most 40\% for MIPS7.
Figure \ref{fig:line correlation} shows the comparison between the $3.3\,\mu$m and $11.3\,\mu$m PAH line flux and prediction from the models.\\
}
\indent{
The $3.3\,\mu$m line flux value for MIPS 3419 is under $2\,\sigma$ hence, it has been considered as an upper limit.
For comparison, we plot observations of two star-formation dominated local
galaxies (Arp 220 and Mrk 273) galaxies and 2 AGN (Mrk 231, IRAS 05189-2425) taken from \citet{imanishi_2000} and \citet{Armus_2007} for $L_{3.3\,\mu m}$ and $L_{11.3\mu m}$ respectively.
The classification of Arp 220 and Mrk 273 as star-formation dominated galaxies and of Mrk 231 and IRAS 05189-2425 as AGN are based on the mid-infrared diagram shown in \citet[see their Figure 8]{Armus_2007}.
We note that the $3.3\,\mu$m line flux derived by \citet{imanishi_2000} includes only the nucleus while \citet{Armus_2007} measure the $11.3\,\mu$m feature in a larger aperture which encompasses most of the galaxy.
We apply an aperture correction to the $3.3\,\mu$m line flux which is the ratio of the $\sim3-4\,\mu$m continuum of the whole galaxy obtained using broad band observations reported in 
the NASA/IPAC Extragalactic Database (NED) and the $3-4\,\mu$m continuum in the spectra of \citet{imanishi_2000}.
This aperture correction relies on the assumption that the nucleus and the whole galaxy have the same spectrum which may not be the case for AGN-dominated nuclei.
In fact, we note that this aperture corrections would provide only a lower limit to $L_{3.3\,\mu m}$ due to the low $3.3\,\mu$m equivalent widths observed in obscured AGN.\\
}
\indent{
The values of $L_{3.3\,\mu m}$ inferred for 5 out of 6 galaxies (MIPS 4, 5, 6, 7 and, 3419) are consistent with the one predicted by the \citet{li_2001} model from the measured $L_{11.3\,\mu m}$.
This agreement supports the hypothesis that the IRAC excess observed in these galaxies is due to the $3.3 \,\mu$m PAH feature.
This true also for MIPS 6 and MIPS 7 suggesting that their potential AGN contribution (see Section \ref{sec:AGN}) does not dominate the IRAC flux excess.
For MIPS 5581, the inferred $3.3\,\mu$m PAH line is over-estimated by a factor of $\sim 1.5-4$ with respect to model expectations from the $L_{11.3\,\mu m}$.
This can be explained by the presence of a significant NIR continuum as also revealed by the color diagnostic presented in Section \ref{sec:cont}.
The assumption that the NIR continuum is negligible compared to the line leads to an overestimation of the inferred $3.3 \,\mu$m PAH line.\\
}
\indent{
The presence of strong PAH features in the IRS spectra of each galaxy of the sample, 
except for MIPS 3419, indicates that they must be star-formation dominated.
The  $L_{3.3 \mu {\rm m}}/L_{IR}$ ratio in local star-forming galaxies has been found to be $\thicksim 10^{-3}$ \citep{mouri_1990} with a factor of $2-3$ scatter.
For MIPS 4, 5, 6, 7, and 5581 we find $L_{3.3 \mu {\rm m}}/L_{IR}\thicksim3\times10^{-3}$ which is, within the uncertainties of the relation, consistent with the local ratio.
For MIPS 3419, the upper limit found for the $3.3\,\mu$m PAH line flux is still consistent with the local ratio of $L_{3.3 \mu m}/L_{IR}$.\\
} 
\indent{
The agreement of the $L_{3.3 \mu m}/L_{11.3 \mu m}$ ratio with standard dust models and of the $L_{3.3 \mu {\rm m}}/L_{IR}$ ratio with local observations favors the hypothesis that the IRAC excess is effectively due to the $3.3\,\mu$m PAH emission.
We also note that even in the case where the IRAC excess would be due to the combined emission of a continuum and the $3.3\,\mu$m PAH, the study of nearby star-forming galaxies by \citet{lu_2003} showed that both emissions probably originate from the same carriers. 
Indeed, they find that the $7.7\,\mu$m PAH equivalent width does not vary in their sample of galaxies. 
This trend was not corroborated by \citet{flagey_2006} but the latter study remains limited to the diffuse ISM of the Milky Way.
In light of \citet{lu_2003} results the information carried by the $3.3\,\mu$m PAH line flux alone or by the continuum and the $3.3\,\mu$m PAH complex would be equivalent.\\
}
\section{Conclusion}
\label{sec: conclusion}
\indent{
We have investigated the presence of an excess in the observed IRAC flux densities
through analysis of the multi-wavelength photometry and spectroscopy of 6 galaxies in the redshift range $0.5<z<1.8$.
Using a fit to the optical and NIR photometry, we determined the stellar continuum of each source with PEGASE.2.
The difference between the stellar continuum and the IRAC flux in the passband expected to harbor the $3.3\,\mu$m PAH feature is computed and analyzed.
For 5 galaxies we found a significant IRAC excess. 
For the other galaxy, we were only able to determine an upper limit.\\
}
\indent{
We investigated the possibility that the measured IRAC excess could be explained by the presence of an obscured AGN using the SED of NGC 1068.
For 1 galaxy (MIPS 6) we find evidence for a possible contamination by an obscured AGN.
For 4 galaxies (MIPS 4, 5, 7, and 5581), no evidence was found that the IRAC excess could result from the presence of an obscured AGN.\\
}
\indent{
We have investigated the origin of the IRAC excess as potentially due to the presence of the $3.3\,\mu$m PAH feature and/or a warm dust continuum.
Local observations \citep[][]{flagey_2006,lu_2003} find evidence for a warm dust continuum of unknown origin which could dominate the IRAC excess.
The NIR colors of the IRAC excess cannot be reproduced by such a continuum unless one assumes a lower temperature.
However, the $3.3\,\mu$m PAH line flux inferred directly from the IRAC excess is consistent with standard dust models.
Indeed, we find that the inferred $3.3\,\mu$m PAH line flux is compatible with the $11.3\,\mu$m PAH line flux measured in the IRS spectrum. 
This suggests that the IRAC excess is effectively due to the $3.3\,\mu$m PAH feature.\\
}
\indent{
However, we note that even if the IRAC excess would arise from the combination of the $3.3\,\mu$m PAH line flux and a warm dust continuum such as that
found in local star-forming galaxies by \citet{lu_2003}, both originate from the same carriers, as indicated by a constant PAH/continuum flux ratio.
In that case, as well as in the case where the $3.3\,\mu$m PAH line flux alone dominates the IRAC excess, we conclude that this emission provides a powerful diagnostic for measuring dusty star formation rates in galaxies at $z>3$ using the \textit{JWST}.
We find that $\nu L_{\nu,\,3.6\,\mu m}^{restframe}/L_{IR} $ is nearby constant for all 5 galaxies, i.e $(1.0\pm0.3)\times 10^{-3}$ with $L_{IR}$ spanning the range [$1.5\times10^{11}\,-\,8\times10^{11}$].\\
}

\acknowledgements{
We thank Stephanie Juneau for revisiting our analysis of the optical spectra of these galaxies.
We thank Masa Imanishi and Aaron Steffen for useful discussions.
AP acknowledges support provided by NASA through the \textit{Spitzer Space Telescope} Fellowship Program, through a contract issued by the Jet Propulsion Laboratory, California Institute of Technology under a contract with NASA.
This work is based on observations made with the \textit{Spitzer Space Telescope}, which is operated by the Jet Propulsion Laboratory, California Institute of Technology under a contract with NASA. 
Support for this work was provided by NASA through an award issued by JPL/Caltech.
This research has made use of the NASA/IPAC Extragalactic Database (NED) which is operated by the Jet Propulsion Laboratory, California Institute of Technology, under contract with the National Aeronautics and Space Administration.
SMART was developed by the IRS Team at Cornell University and is available through the Spitzer Science Center at Caltech.
}
\bibliographystyle{apj}



\begin{deluxetable}{cccccccccccc}
\tabletypesize{\scriptsize}
\tablecaption{\label{tab:sample}Optical and near-infrared photometry of our sample}
\rotate
\tablewidth{0pt}
\tablecolumns{3}
\tablehead{
\multicolumn{4}{c}{Objects}\\
\colhead{Id} &\colhead{z} &\colhead{RA} &\colhead{Dec} & \colhead{{\textit B}} & \colhead{{\textit V}}  &\colhead{ {\textit i}} & \colhead{{\textit z}} & \colhead{3.6 $\mu$m} & \colhead{4.5 $\mu$m} & \colhead{5.8 $\mu$m} & \colhead{8.0 $\mu$m}  \\
&  & \multicolumn{2}{c}{\scriptsize{(J2000)}} &{\scriptsize ($\rm{\mu Jy}$) }&{\scriptsize    ($\rm{\mu Jy}$)  }&{\scriptsize    ($\rm{\mu Jy}$) }&{\scriptsize    ($\rm{\mu Jy}$) }&{\scriptsize    ($\rm{\mu Jy}$) }&{\scriptsize    ($\rm{\mu Jy}$) }&{\scriptsize    ($\rm{\mu Jy}$)} &{\scriptsize    ($\rm{\mu Jy}$)} \\ 
}
\startdata
MIPS4 & 0.638 &189.01355  & 62.18634 & $3.9\pm0.2$   &  $11.9\pm0.5$ &     $27.2\pm1.2$  &     $35.8\pm1.6$  & $129.9\pm6.5$  &  $89.1\pm4.4$  &  $107.9\pm5.4$   &    $98.7\pm4.9$ \\
MIPS5 & 0.641 & 189.39383  & 62.28978 & $1.0\pm0.06$ &  $5.0\pm0.2$  & $15.4\pm0.7$ &  $24.3\pm1.0$ &  $167.9\pm8.4$ & $113.9\pm5.7$ & $124.9\pm6.2$ & $100.9\pm5.0$ \\
MIPS6  & 0.639 & 189.09367 & 62.26231 & $5.3\pm0.2$ & $12.1\pm0.5$ & $24.5\pm1.1$ & $30.4\pm1.3$ &  $67.3\pm3.3$ & $46.5\pm2.3$ & $56.4\pm2.8$ & $65.8\pm3.3$ \\
MIPS7  & 0.792 & 189.23306 & 62.13559 & $0.3\pm0.06$ &  $2.0\pm0.09$ & $8.5\pm0.3$ & $13.5\pm0.6$ & $112.9\pm5.6$ &  $80.8\pm4.0$ & $77.9\pm3.9$ & $80.9\pm4.0$ \\
MIPS3419 & 1.70 & 189.17568  & 62.28963 & $0.02\pm0.01$ & $0.06\pm0.01$ &  $0.35\pm0.03$ & $0.55\pm0.04$ & $33.5\pm1.6$ & $38.5\pm1.9$ & $34.3\pm1.7$ & $23.5\pm1.2$ \\
MIPS5581 & 0.839 & 189.28491  & 62.25418 & $2.4\pm0.1$ & $3.9\pm0.2$  & $7.9\pm0.3$ & $9.9\pm0.4$ & $27.3\pm1.3$ & $18.7\pm0.9$ & $20.0\pm1.0$ & $14.7\pm0.8$ \\
\enddata
\tablecomments{Catalogs have been cross-correlated using a matching radius of 0.5 \arcsec}
\end{deluxetable}

\begin{deluxetable}{cccccccc}
\tabletypesize{\scriptsize}
\tablecaption{\label{tab:radio} Infrared and radio flux densities of the sample}
\tablewidth{0pt}
\tablecolumns{2}
\tablehead{
\colhead{Id object}  & \colhead{$S_{16\,\mu m}$} & \colhead{$S_{24\,\mu m}$}  & \colhead{$S_{70\,\mu m}$} & \colhead{$S_{1.4\,GHz}^{obs}$} & \colhead{$S_{1.4\,GHz}^{predicted}$} & \colhead{$L_{1.4\rm{GHz}}^{rest}$} & \colhead{$Log(L_{IR})$} \\
 & \colhead{\tiny{($\rm{\mu Jy}$)}} &\colhead{\tiny{($\rm{\mu Jy}$)}} & \colhead{\tiny{($\rm{mJy}$)}}&\colhead{\tiny{($\rm{\mu Jy}$)}} & \colhead{\tiny{($\rm{\mu Jy}$)}}& \colhead{\tiny{($\times10^{23}\rm{W\,Hz^{-1}}$)}} & \colhead{\tiny{$\rm{(L_{\odot})}$}} \\
}
\startdata
MIPS 4 & $777\pm20$& $1221\pm12$ & $11.0\pm 0.66$ & $161\pm12$ & 126 & 2.4 & $11.90\pm0.05$ \\
MIPS 5 & $575\pm11$& $750\pm7$ &$5.6\pm0.67$ &$91\pm14$ & 79 & 1.4 &$11.75\pm0.04$\\
MIPS 6 & $398\pm5$& $721\pm7$ &$11.0\pm0.66$ &$64\pm16$& 75 & 1.0 &$11.70\pm0.04$\\
MIPS 7 & $582\pm10$& $832\pm8$ &$14.0\pm0.7$&$104\pm11$& 81& 2.7 &$11.89\pm0.05$\\
MIPS 3419 &$54\pm7$& $113\pm6$ & $<3.0$ & $<25$ & 19 & $<4.0$ &$11.76\pm0.10$\\
MIPS 5581 &$194\pm5$ & $201\pm6$ & $<3.0$& $16\pm5$& 17 & 0.6 &$11.22\pm0.09$\\
\enddata
\end{deluxetable}

\begin{deluxetable}{cccc}
\tabletypesize{\scriptsize}
\tablecaption{\label{tab:peg para} PEGASE.2 template parameters}
\tablewidth{0pt}
\tablecolumns{2}
\tablehead{
\colhead{Type} & \colhead{$\nu$} & \colhead{Infall ($t_{c}$)} & \colhead{Gal Winds} \\
\colhead{} & \colhead{\tiny {($\rm{Myr}$)}}&\colhead{ \tiny{($\rm{Myr}$)} }& \colhead{ \tiny{($\rm{Myr}$)} } \\
}
\startdata
E & 100 & 100 & 3000 \\
S0 & 500 & 100 & 5000 \\
Sa & 1500 & 500 & \dots \\
Sb & 2500 & 1000 &  \dots \\
Sbc & 5000 & 1000 &   \dots \\
Sc & 10000 & 2000 &   \dots \\
Sd & 20000 & 2000 &   \dots \\
Irr & 20000 & 5000 &   \dots \\
\enddata
\tablecomments{PEGASE.2 scenarios used as template parameters. $SFR=\nu^{-1}\times M_{gas}$ and gas infall is simulated as $f(t)=\frac{exp(-t/t_{c})}{t_{c}}$. $\nu$ is effectively the ratio between the star formation time scale and the star formation efficiency. The Initial Mass Function used in our scenarios is taken from \citet{rana_1992}.}
\end{deluxetable}
\begin{deluxetable}{cccccc}
\tabletypesize{\scriptsize}
\tablecaption{\label{tab:fit peg} Fit parameters}
\tablewidth{0pt}
\tablecolumns{2}
\tablehead{
\colhead{Id object} & \colhead{Type} & \colhead{Age} & \colhead{Stellar Mass} & \colhead{E(\textit{B-V})} & \colhead{$\chi^{2}$} \\
& & \colhead{\tiny{ ($\rm{Gyr}$)}}&\colhead{\tiny{($\rm{M_{\odot}}$)}} \\
}
\startdata
MIPS 4  & Sbc & 7 & $1.8\times10^{11}$ & 0.18 & 5.15\\
MIPS 5 &  Sa & 6 & $2.8\times10^{11}$ & 0.72 & 9.75\\
MIPS 6 &  Sbc & 5 & $7.3\times10^{10}$ & 0.03 & 8.68\\
MIPS 7 & E & 3 & $2.3\times10^{11}$ & 0.88 & 15.75\\
MIPS 3419 & Sd & 3 & $2.6\times10^{11}$ & 2.73 & 6.48\\
MIPS 5581 & Sb & 3 & $3.4\times10^{10}$ & 0.03 & 22.26 \\
\enddata
\tablecomments{$\chi^{2}=\sum_{i=1}^N\frac{(x_{i}-\overline x_{i})^{2}}{\sigma_{i}^{2}}$ where N is the number of passbands fit to the model.}
\end{deluxetable}

\begin{deluxetable}{ccccccc}
\tabletypesize{\scriptsize}
\tablecaption{\label{tab:pah line} Inferred PAH properties}
\tablewidth{0pt}
\tablecolumns{2}
\tablehead{
\colhead{Id object} & \colhead{IRAC excess} &\colhead{$3.3 \mu$m line flux} & \colhead{$11.3 \mu$m line flux} & \colhead{$EW_{restframe}$} & \colhead{SFR} & \colhead{$L_{3.3\,\mu m}/L_{IR}$} \\
 & \colhead{\tiny{($\rm{\times10^{-28} erg\ s^{-1} cm^{-2} Hz^{-1} }$)} }   & \colhead{\tiny{($\rm{\times 10^{-22} W cm^{-2}}$)}} & \colhead{\tiny{($\rm{\times 10^{-22} W cm^{-2}}$)}} & \colhead{\tiny{($\rm{nm}$)}} & \colhead{\tiny{$\rm{(M_{\odot}\,Yr^{-1}}$)}} & \colhead{\tiny{($\rm{\times 10^{-3}}$)}} \\
}
\startdata
MIPS 4  & $3.7\pm0.7$ & $4.92\pm1.45$&$4.67\pm0.58$& 227&137.9&$2.7\pm0.8$\\
MIPS 5 & $3.2\pm0.8$ &  $3.66\pm0.75$ &$5.87\pm0.68$& 58  &97.8&$2.81\pm0.8$\\
MIPS 6 & $2.0\pm0.3$ &  $2.75\pm0.30$&$3.69\pm0.69$& 119 &86.3 &$2.48\pm0.5$\\
MIPS 7 & $1.4\pm0.5$ & $1.34\pm0.62$&$4.16\pm1.66$& 45  &136.0&$1.29\pm0.6$\\
MIPS 3419 & $0.0\pm0.1$ & $0.00\pm0.11$&$<0.18$& $<11$ &99.5&$<1.9$\\
MIPS 5581 & $ 0.6\pm0.1$ & $0.67\pm0.15$&$0.17\pm0.9$& 106 &28.5&$3.5\pm1.0$\\
\enddata
\tablecomments{$3.3\,\mu$m line fluxes have been calculated with the original PEGASE.2 fit and errors have been calculated using the Monte Carlo approach. The SFR is derived from $L_{IR}$ using Equation \ref{eq:sfr lir}}
\end{deluxetable}

\begin{deluxetable}{crlrlc}
\tabletypesize{\scriptsize}
\tablecaption{\label{tab:X-ray}X-Ray properties of the sample}
\tablewidth{0pt}
\tablecolumns{4}
\tablehead{
\colhead{Id object} & \multicolumn{2}{c}{Soft X-ray} &  \multicolumn{2}{c}{Hard X-ray}  & \colhead{Photon index}\\
&\multicolumn{2}{c}{(0.5-2 \rm{keV})}&\multicolumn{2}{c}{(2-8 \rm{keV})}&\colhead{($\Gamma$)} \\
& \multicolumn{1}{c}{Obs} & \multicolumn{1}{c}{(Model)} & \multicolumn{1}{c}{Obs} & \multicolumn{1}{c}{(Model)} &\\
& \multicolumn{4}{c}{\tiny{[$\rm{\times 10^{-16} \rm{erg\,s^{-1}\,cm^{-2}}}$]} } \\
}
\startdata
4  & 1.42 &(1.86)& $<3.21$ &(2.33)& $>1.45$\\
5 & $<0.76$ &(1.31)& $<7.49$ &(1.68)& \dots\\
6 &  0.65 &(1.22) & 2.54&(1.48) & 1.06 \\
7 & 1.06 & (1.07)&$<2.73$ &(1.32)& $>1.35$\\
5581 & $<0.29$&(0.24) & $<1.55$&(0.33) & \dots\\
\enddata
\end{deluxetable}

\end{document}